\documentclass[prc,showpacs,nofootinbib,preprint]{revtex4}
\usepackage[colorlinks]{hyperref}
\usepackage{epsfig,graphics,color}
\usepackage{subfigure}
\usepackage{amssymb}
\usepackage{amsmath}
\usepackage{pdfsync}
\definecolor{mono}{rgb}{0.65, 0.65, 0.65} 

\newcommand{\dd}{\mathrm{d}}

\def\tr{{\rm tr} \,}

\def\pslash{p \hspace{-1.7mm}/}

\def\Dslash{D \hspace{-2.7mm}/ \;}

\def\w2{\tilde w^2}
\def\ws2{1}

\newlength{\lslash}

\begin{document}
\title{A framework for the chiral extrapolation of the \\charmed baryon ground-state masses}
\author{Yonggoo Heo$^1$, Xiao-Yu Guo$^2$ and Matthias F.M. Lutz$^{2,3}$}
\affiliation{$^1$ Suranaree University of Technology, Nakhon Ratchasima, 30000, Thailand}
\affiliation{$^2$ GSI Helmholtzzentrum f\"ur Schwerionenforschung GmbH,\\
Planck Str. 1, 64291 Darmstadt, Germany}
\affiliation{$^3$ Technische Universit\"at Darmstadt, D-64289 Darmstadt, Germany}
\date{\today}
\begin{abstract}
We consider the chiral Lagrangian for charmed baryon fields with $J^P =\frac{1}{2}^+$ or 
$J^P =\frac{3}{2}^+$ quantum numbers. A chiral expansion framework for the baryon ground state masses is worked out to N$^3$LO 
as to compute their dependence on the up, down and strange quark masses for finite box QCD lattice simulations. It is formulated in terms of on-shell meson and baryon masses. The convergence of such a scheme is illustrated with physical masses as taken from the PDG. 
The counter terms relevant at N$^3$LO are correlated systematically by large-$N_c$ sum rules to leading and subleading order in a manner that keeps the 
renormalization scale invariance of the approach.
\end{abstract}

\maketitle
\tableofcontents

\section{Introduction}

QCD lattice simulations offer the opportunity to determine low-energy parameters of the
chiral Lagrangian. Since the simulations are performed also at quark masses distinct to those needed to reproduce the physical hadron
masses new information is generated that may help to determine so far unknown low-energy constants. 

Such programs have already been successfully  set up for the masses of baryons and  mesons in their ground states with $J^P = \frac{1}{2}^+, \frac{3}{2}^+$ and 
$J^P = 0^-, 1^-$ quantum numbers \cite{Lutz:2018cqo,Guo:2018kno,Bavontaweepanya:2018yds}. Corresponding sets of low-energy parameters to be used in flavor SU(3) chiral Lagrangians 
were established from the available lattice data on such hadron masses \cite{Lutz:2018cqo,Guo:2018kno}. 

The purpose of the present work is to establish a corresponding framework for the masses of charmed baryons, which can then eventually  be applied to the current QCD lattice data. Given the rather scarce data set  that is provided so far on the charmed baryon masses \cite{Liu:2009jc,Bali:2012ua,Briceno:2012wt,Alexandrou:2012xk,Namekawa:2013vu,Perez-Rubio:2013oha} it is important to derive additional constraints from QCD that will guide a fit of the low-energy constants to such data. 
An important first step in this direction are the recent works \cite{Lutz:2014jja,Heo:2018mur}, 
in which all counter terms that turn relevant in a chiral expansion of the charmed baryon masses are constructed and correlated by the heavy-quark spin symmetry and sum rules derived from large-$N_c$ QCD. Here we complement these results by deriving explicit expressions for the various contributions to the baryon masses that arise at next-to-next-to-next-to leading order (N$^3$LO). 

As was argued in our previous works  \cite{Lutz:2018cqo,Guo:2018kno,Bavontaweepanya:2018yds} a chiral expansion around the flavor SU(3) limit of QCD in terms of 
bare meson and baryon masses is not convergent for the physical up, down and strange quark masses. Any attempt to apply such a conventional expansion strategy to the QCD lattice data set is futile and should be abandoned \cite{Jiang:2014ena}. Instead, it was demonstrated that a reformulated expansion scheme that uses the on-shell meson and baryon masses appears to have a significantly larger convergence domain, that is applicable to the 
physical masses for up, down and strange quarks \cite{Lutz:2018cqo,Guo:2018kno}. In this work the details of such an approach for the charmed baryon masses are presented. In particular its convergence properties are illustrated at the hand of a chiral decomposition of the bubble loop contributions at physical meson and baryon masses. This is supplemented by the derivation of 
additional sum rules for the counter terms that arise from the condition of renormalization scale invariance. 
Given the results of this work an application to the data set from lattice QCD group is feasible. This should eventually lead to a faithful set of low-energy constants. 

The work is organized as follows. In section II the relevant parts of the chiral Lagrangian are collected. All expressions required at N$^3$LO are detailed in section III. In section IV the various large-$N_c$ sum rules are studied at the one-loop level. The paper continues with a convergence study in section V and VI of the one-loop bubble contributions as decomposed into their chiral moments. A short summary  given with section VII. 
The appendix provides  a glossary for our notations and conventions.

\newpage 

\section{Chiral Lagrangian with charmed baryon fields} \label{section:chiral-lagrangian}

The chiral Lagrangian is a reliable tool, once it is combined with appropriate
counting rules leading to a systematic approximation strategy. In the following we recall 
the leading order (LO) terms \cite{Yan:1992gz,Cho:1992gg,Lutz:2014jja}. It is convenient to decompose the fields into their isospin multiplets with
\begin{eqnarray}
&& \Phi = \tau \cdot \pi (140)
+ \alpha^\dagger \cdot  K (494) +  K^\dagger(494) \cdot \alpha
+ \eta(547)\,\lambda_8\,,
\nonumber\\
&&  \sqrt{2}\,B_{[\bar 3]}  = {\textstyle{1\over \sqrt{2}}}\,\alpha^\dagger \cdot \Xi_c(2470)
- {\textstyle{1\over \sqrt{2}}}\,\Xi_c^{T}(2470)\cdot \alpha
+  i\,\tau_2\,\Lambda_c(2284) \,,
\nonumber\\
&& \sqrt{2}\,B_{[6]} = {\textstyle{1\over \sqrt{2}}}\,\alpha^\dagger \cdot \Xi'_c(2580)
+ {\textstyle{1\over \sqrt{2}}}\,{\Xi'}_{\!\!c}^{T}(2580)\cdot \alpha
+ \Sigma_c(2455) \cdot \tau \,i\,\tau_2 
\nonumber\\
&& \qquad \quad+  {\textstyle{\sqrt{2}\over 3}}\, \big(1-\sqrt{3}\,\lambda_8 \big)\,\Omega_c(2704)  \,,
\nonumber\\
&& \sqrt{2}\, B^\mu_{[6]} = {\textstyle{1\over \sqrt{2}}}\,\alpha^\dagger \cdot \Xi^{\mu}_c(2645)
+ {\textstyle{1\over \sqrt{2}}}\,\Xi_c^{T,\mu}(2645)\cdot \alpha
+ \Sigma^\mu_c(2520) \cdot \tau \,i\,\tau_2 
\nonumber\\
&& \qquad \quad+  {\textstyle{\sqrt{2}\over 3}}\, \big(1-\sqrt{3}\,\lambda_8 \big)\,\Omega^{\mu}_c(2770)  \,,
\nonumber\\
&& \alpha^\dagger = {\textstyle{1\over \sqrt{2}}}\,(\lambda_4+i\,\lambda_5 ,\lambda_6+i\,\lambda_7 )\,,\qquad
\tau = (\lambda_1,\lambda_2, \lambda_3)\,,
\label{def-states}
\end{eqnarray}
where the matrices $\lambda_i$ are the standard Gell-Mann generators of the SU(3) algebra.
The numbers in the brackets recall the approximate masses of the particles in units of
MeV. Note that we do not consider the $\eta'$ meson as an active degree of freedom in our current study \cite{Terschlusen:2012xw,Guo:2015xva}

It should be noted that $\Xi_c$ and $\Xi'_c$ have the same quantum numbers and therefore 
a mixing of the two fields needs to be considered \cite{Okubo:1963fa}. 
We introduce a $\Xi_c-\Xi_c'$ mixing angle $\epsilon$ by 
\begin{eqnarray}
&& \Xi_c =\bar \Xi_c \, \cos \epsilon +\bar \Xi'_c \,\sin \epsilon \,, \qquad \qquad \qquad \Xi'_c =\bar \Xi'_c\,  \cos \epsilon  -\bar \Xi_c\, \sin \epsilon\,,
\nonumber\\
&& \qquad{\rm with} \qquad \Sigma_{\Xi_c \Xi'_c} =\frac{1}{2}\,\Big(\Sigma_{\Xi'_c} - \Sigma_{\Xi_c} \Big)\,\tan (2\,\epsilon) \,,
\end{eqnarray}
where the physical fields are denoted by $\bar \Xi_c$ and $\bar \Xi'_c$. The off-diagonal self energy $\Sigma_{\Xi_c \,\Xi'_c} \neq 0$ reflects the fact that the fields $\Xi_c$ and $\Xi'_c$ are unphysical. Only for the physical fields  $\bar \Xi_c$ and $\bar \Xi'_c$ we expect their corresponding off-diagonal self energy $\Sigma_{\bar \Xi_c \,\bar \Xi'_c} \to 0$ to vanish for
on-shell conditions.

There are the kinetic terms 
\allowdisplaybreaks[1]
\begin{eqnarray}
&& \mathcal{L}^{(1)} \!= \tr \bar B_{[6]} \big(\gamma^\mu\, i\,D_\mu  - M^{1/2}_{[6]} \big)\,B_{[6]}
- \mathrm{tr}\,\Big(\bar{B}_{[6]}^\mu \, \big(\big[i\,\Dslash\,
-M^{3/2}_{[6]}\big]\,g_{\mu\nu} -i\,(\gamma_\mu\, D_\nu 
 + \gamma_\nu \,D_\mu)
\nonumber\\
&& \qquad \;+\, \gamma_\mu\,\big[i\,\Dslash + M^{3/2}_{[6]}\big]\,\gamma_\nu \big)\,B_{[6]}^\nu\Big)
+\tr \bar B_{[\bar 3]} \big(\gamma^\mu\,i\, D_\mu  -M^{1/2}_{[\bar 3]} \big)\,B_{[\bar 3]} 
\nonumber \\
&&  \qquad \;+\,F_{[66]}\,{\rm tr}\,\bar B_{[6]}\,\gamma^\mu\,\gamma_5\,i\,U_\mu\,B_{[6]}
+ F_{[\bar 3\bar 3]}\,{\rm tr}\,\bar B_{[\bar 3]}\,\gamma^\mu\,\gamma_5\,i\,U_\mu\,B_{[\bar 3]}
\nonumber\\
&&  \qquad \;+\, F_{[\bar 36]}\,{\rm tr}\,\Big( \bar B_{[6]}\,\gamma^\mu\,\gamma_5\,i\,U_\mu\,B_{[\bar 3]} + {\rm h.c.}\Big)
\nonumber\\
&&  \qquad \;+\, C_{[66]}\,{\rm tr}\,\Big( \bar B_{[6]}^\mu\,i\,U_\mu\,B^{\phantom{\mu}}_{[6]} + {\rm h.c.} \Big)
 + C_{[\bar 36]}\,{\rm tr}\,\Big( \bar B_{[6]}^\mu\,i\,U_\mu\,B^{\phantom{\mu}}_{[\bar 3]} + {\rm h.c.}\Big)
\nonumber\\
&&  \qquad \;-\, H_{[66]}\,{\rm tr}\,\bar B_{[6]}^\alpha\,g_{\alpha \beta}\,\gamma^\mu\,\gamma_5\,i\,U_\mu\,B_{[6]}^\beta \,,
\nonumber\\
\nonumber\\
&& U_\mu = {\textstyle{1\over 2}}\,u^\dagger \, \big(
\partial_\mu \,e^{i\,\frac{\Phi}{f}} \big)\, u^\dagger
-{\textstyle{i\over 2}}\,u^\dagger \,(v_\mu+ a_\mu)\, u
+{\textstyle{i\over 2}}\,u \,(v_\mu-a_\mu)\, u^\dagger\;, \qquad \qquad   u = e^{i\,\frac{\Phi}{2\,f}}  \,,
\nonumber\\
&&  D_\mu \, B \;\,= \partial_\mu B +  \Gamma_{\mu}\, B + B\,\Gamma^T_{\mu} \,, \qquad \qquad \qquad
\nonumber\\
&&\Gamma_\mu ={\textstyle{1\over 2}}\,u^\dagger \,\big[\partial_\mu -i\,(v_\mu + a_\mu) \big] \,u
+{\textstyle{1\over 2}}\, u \,
\big[\partial_\mu -i\,(v_\mu - a_\mu)\big] \,u^\dagger\,,  
\label{def-L1}
\end{eqnarray}
and 6 structures which parameterize the three-point interactions of the Goldstone bosons with the charmed baryon 
fields \cite{Yan:1992gz,Cho:1992gg}. From the kinetic terms one can read off the Weinberg-Tomozawa interaction terms on which the 
coupled-channel computation of \cite{Lutz:2003jw} rests. It follows upon an expansion of the kinetic terms in powers 
of the Goldstone boson fields. At leading order in a chiral expansion, the bare masses 
$M^{1/2}_{[6]}$, $M^{3/2}_{[6]}$ and $M^{1/2}_{[\bar 3]}$ may be identified with the flavor average of the sextet 
and anti-triplet baryon masses. Note the classical vector and axial-vector source functions $v_\mu$ and $a_\mu$ of QCD in (\ref{def-L1}) were instrumental in the derivation of our large-$N_c$ sum rules \cite{Lutz:2014jja,Heo:2018mur}.

We proceed with the terms at next-to-leading order (NLO) where there are symmetry conserving 
and symmetry breaking terms \cite{Lutz:2014jja,Heo:2018mur}.  We recall the 7 symmetry breaking counter terms 
\allowdisplaybreaks[1]
\begin{eqnarray}
&&\mathcal{L}^{(2)}_\chi =  b_{1,[\bar 3\bar 3]}\,{\rm tr}\,\big( \bar B_{[\bar 3]}\,B_{[\bar 3]}\big)\,{\rm tr}\,\big(\chi_+\big)
+b_{2,[\bar 3\bar 3]}\,{\rm tr}\,\big( \bar B_{[\bar 3]}\,\chi_+\,B_{[\bar 3]}\big)
+b_{1,[\bar 3 6]}\,{\rm tr}\,\big( \bar B_{[6]}\,\chi_+\,B_{[\bar 3]} + {\rm h.c.}\big)
\nonumber\\
&& \qquad \;\,+\, 
b_{1,[66]}\,{\rm tr}\,\big( \bar B_{[6]}\,B_{[6]}\big)\,{\rm tr}\,\big(\chi_+\big)
+ b_{2,[66]}\,{\rm tr}\,\big( \bar B_{[6]}\,\chi_+\,B_{[6]}\,\big)
\nonumber\\
&& \qquad \;\,-\, d_{1,[66]}\,{\rm tr}\,\big(  g_{\mu \nu}\,\bar B_{[6]}^\mu\,B_{[6]}^\nu\big)\,{\rm tr}\,\big(\chi_+\big)
- d_{2,[66]}\,{\rm tr}\,\big( g_{\mu \nu}\,\bar B_{[6]}^\mu\,\chi_+\,B_{[6]}^\nu\,\big)\,,
\nonumber\\
&& \chi_+ = {\textstyle{1\over 2}}\, \big(
u \,\chi_0 \,u
+ u^\dagger \,\chi_0 \,u^\dagger \big)\,,
\label{def-L2-chi}
\end{eqnarray}
with $\chi_0 = 2\,B_0 \,{\rm diag }(m ,m, m_s)  $ proportional to the quark-mass matrix. We do not consider
isospin violating effects in this work. The low-energy constants of (\ref{def-L2-chi}) imply a linear quark-mass dependence for the charmed baryon masses with 
\begin{eqnarray}
&& M^{(2)}_{\Xi_c} -M^{(2)}_{\Lambda_c} = -B_0\, (m_s-m)\,b_{2,[\bar 3 \bar 3]} \,, \qquad 
\nonumber\\
&& M^{(2)}_{\Xi'_c} -M^{(2)}_{\Sigma_c} = -B_0\, (m_s-m)\,b_{2,[66]} \,,\qquad \qquad  M^{(2)}_{\Omega_c} -M^{(2)}_{\Sigma_c} = -2\,B_0\, (m_s-m)\,b_{2,[66]} \,,
\nonumber\\
&& M^{(2)}_{\Xi^*_c} -M^{(2)}_{\Sigma^*_c} = -B_0\, (m_s-m)\,d_{2,[66]} \,,\qquad \qquad  M^{(2)}_{\Omega^*_c} -M^{(2)}_{\Sigma^*_c} = -2\,B_0\, (m_s-m)\,d_{2,[66]} \,,
\nonumber\\ 
&& M^{(2)}_{\Lambda_c} + 2\, M^{(2)}_{\Xi_c} =- 2\,B_0\, (2\,m +m_s)\,\big(3\, b_{1,[\bar 3 \bar 3]}  + b_{2,[\bar 3 \bar 3]} \big)\,,
 \nonumber\\
&& M^{(2)}_{\Omega_c} + 2\, M^{(2)}_{\Xi'_c} + 3\, M^{(2)}_{\Sigma_c} =- 4\,B_0\, (2\,m +m_s)\,\big(3\, b_{1,[66]}  + b_{2,[66]}  \big)\,,
 \nonumber\\
&& M^{(2)}_{\Omega^*_c} + 2\, M^{(2)}_{\Xi^*_c} + 3\, M^{(2)}_{\Sigma^*_c} =- 4\,B_0\, (2\,m +m_s)\,\big(3\, d_{1,[66]}  + d_{2,[66]} \big)\,,
\nonumber \\
&& \qquad \qquad M^{(2)}_{\Xi_c \Xi'_c} =  B_0\,(m_s-m)\, b_{1,[\bar 3 6]}\,,
\label{def-Sigma-Q2}
\end{eqnarray}
where the upper index 2 in $M^{(2)}_B$ projects the mass of the baryon of type $B$ on its chiral order $Q^2$ in this case. 

A complete list of chiral symmetry conserving $Q^2$ counter terms, relevant for the calculation of the charm baryon masses at N$^3$LO, 
was given in \cite{Lutz:2014jja,Heo:2018mur}. In these works the $Q^2$ counter terms are grouped according
to their Dirac structure. Here we display the scalar and vector terms relevant for our study only
\begin{eqnarray}\label{del-L2}
\mathcal{L}^{(2)}=\mathcal{L}^{(S)} + \mathcal{L}^{(V)}\,,
\end{eqnarray}
with
\allowdisplaybreaks[1]
\begin{eqnarray}
&& \mathcal{L}^{(S)} \!= -g_{0,[\bar 3\bar 3]}^{(S)}\,{\rm tr}\,\big( \bar B_{[\bar 3]}\,B_{[\bar 3]}\big)\,{\rm tr}\,\big( U_\mu\,U^\mu\big)
- g_{D,[\bar 3\bar 3]}^{(S)}\,{\rm tr}\,\big( \bar B_{[\bar 3]}\,\big\{ U_\mu,\,U^\mu\big\}\, B_{[\bar 3]}\big)
\nonumber\\
&& \qquad \;-\, 
g_{0,[66]}^{(S)}\,{\rm tr}\,\big( \bar B_{[6]}\,B_{[6]}\big)\,{\rm tr}\,\big( U_\mu\,U^\mu\big)
- g_{1,[66]}^{(S)}\,{\rm tr}\,\big( \bar B_{[6]}\, U^\mu\,B_{[6]}\, U^T_\mu\big)
\nonumber\\
&& \qquad \;-\, g_{D,[66]}^{(S)}\,{\rm tr}\,\big( \bar B_{[6]}\,\big\{ U_\mu,\,U^\mu\big\}\, B_{[6]}\big)
- g_{D,[\bar 36]}^{(S)}\,{\rm tr}\,\big( \bar B_{[6]}\,\big\{ U_\mu,\,U^\mu\big\}\, B_{[\bar 3]} + {\rm h.c.}\big)
\nonumber\\
&& \qquad \;+\, h_{0,[66]}^{(S)}\,{\rm tr}\,\big( \bar B_{[6]}^\mu\,g_{\mu \nu}\,B_{[6]}^\nu\big)\,{\rm tr}\,\big( U_\alpha\,U^\alpha\big)
+ h_{1,[66]}^{(S)}\,{\rm tr}\,\big( \bar B_{[6]}^\mu\,B_{[6]}^\nu\big)\,{\rm tr}\,\big( U_\mu\,U_\nu\big)
\nonumber\\
&& \qquad \;+\, h_{2,[66]}^{(S)}\,{\rm tr}\,\big( \bar B_{[6]}^\mu\, g_{\mu \nu}\,\big\{ U^\alpha,\,U_\alpha\big\}\, B_{[6]}^\nu\big)
+ h_{3,[66]}^{(S)}\,{\rm tr}\,\big( \bar B_{[6]}^\mu\,\big\{ U_\mu,\,U_\nu\big\}\, B_{[6]}^\nu\big)
\nonumber\\
&& \qquad \;+\, h_{4,[66]}^{(S)}\,{\rm tr}\,\big( \bar B_{[6]}^\mu\,g_{\mu \nu}\,U^\alpha\, B_{[6]}^\nu\,U^T_\alpha \big)
+ \frac{1}{2}\,h_{5,[66]}^{(S)}\,{\rm tr}\,\big( \bar B_{[6]}^\mu\,U_\nu\, B_{[6]}^\nu\,U^T_\mu + \bar B_{[6]}^\mu\,U_\mu\, B_{[6]}^\nu\,U^T_\nu\big)\,,
\nonumber\\
&& \mathcal{L}^{(V)} \!= -\frac{1}{2}\,g_{0,[\bar 3\bar 3]}^{(V)}\,{\rm tr}\,\big( \bar B_{[\bar 3]}\,i\,\gamma^\alpha\,(D^\beta B_{[\bar 3]})\,{\rm tr}\,\big( U_\beta\,U_\alpha\big) + {\rm h.c.} \big)
\nonumber\\
&& \qquad \;-\, \frac{1}{2}\,g_{1,[\bar 3\bar 3]}^{(V)}\,{\rm tr}\,\big( \bar B_{[\bar 3]}\,i\,\gamma^\alpha\,U_\beta\,(D^\beta B_{[\bar 3]})\, U^T_\alpha + \bar B_{[\bar 3]}\,i\,\gamma^\alpha\,U_\alpha\,(D^\beta B_{[\bar 3]})\, U^T_\beta + {\rm h.c.}\big)
\nonumber\\
&& \qquad \;-\,
 \frac{1}{2}\,g_{D,[\bar 3\bar 3]}^{(V)}\,{\rm tr}\,\big( \bar B_{[\bar 3]}\,i\,\gamma^\alpha\,\big\{ U_\alpha,\,U_\beta\big\}\,(D^\beta  B_{[\bar 3]}) + {\rm h.c.} \big)
\nonumber\\
&& \qquad \;-\,  \frac{1}{2}\,g_{D,[\bar 36]}^{(V)}\,{\rm tr}\,\big( \bar B_{[6]}\,i\,\gamma^\alpha\,\big\{ U_\alpha,\,U_\beta\big\}\,(D^\beta  B_{[\bar 3]})
- (D^\beta \bar B_{[6]})\,i\,\gamma^\alpha\,\big\{ U_\alpha,\,U_\beta\big\}\, B_{[\bar 3]}
+ {\rm h.c.} \big)
\nonumber\\
&& \qquad \;-\,  \frac{1}{2}\,g_{0,[66]}^{(V)}\left(\,{\rm tr}\,\big( \bar B_{[6]}\,i\,\gamma^\alpha\,(D^\beta B_{[6]})\big)\,{\rm tr}\,\big( U_\beta\,U_\alpha\big) + {\rm h.c.} \right)
\nonumber\\
&& \qquad \;-\, \frac{1}{4}\,g_{1,[66]}^{(V)}\,{\rm tr}\,\big( \bar B_{[6]}\,i\,\gamma^\alpha\,U_\beta\,(D^\beta B_{[6]})\, U^T_\alpha + \bar B_{[6]}\,i\,\gamma^\alpha\,U_\alpha\,(D^\beta B_{[6]})\, U^T_\beta + {\rm h.c.}\big)
\nonumber\\
&& \qquad \;-\, \frac{1}{2}\,g_{D,[66]}^{(V)}\,{\rm tr}\,\big( \bar B_{[6]}\,i\,\gamma^\alpha\,\big\{ U_\alpha,\,U_\beta\big\}\,(D^\beta  B_{[6]}) + {\rm h.c.}\big)
\nonumber\\
&& \qquad \;+\,\frac{1}{2}\, h_{0,[66]}^{(V)}\,{\rm tr}\,\big( \bar B_{[6]}^\mu\,g_{\mu \nu}\,i\,\gamma^\alpha\,(D^\beta B_{[6]}^\nu)\,{\rm tr}\,\big( U_\alpha\,U_\beta\big) + {\rm h.c.} \big)
\nonumber\\
&& \qquad \;+\,  \frac{1}{4}\,h_{1,[66]}^{(V)}\,{\rm tr}\,\big( \bar B_{[6]}^\mu\,g_{\mu \nu}\, i\,\gamma^\alpha\,U_\beta\,(D^\beta B_{[6]}^\nu)\, U^T_\alpha + \bar B_{[6]}^\mu\,g_{\mu \nu}\,i\,\gamma^\alpha\,U_\alpha\,(D^\beta B_{[6]}^\nu)\, U^T_\beta + {\rm h.c.}\big)
\nonumber\\
&& \qquad \;+\, \frac{1}{2}\,h_{2,[66]}^{(V)}\,{\rm tr}\,\big( \bar B^\mu_{[6]}\,g_{\mu \nu}\,i\,\gamma^\alpha\,\big\{ U_\alpha,\,U_\beta\big\}\,(D^\beta B_{[6]}^\nu ) + {\rm h.c.}\big)  \,,
\label{del-L2S-L2V}
\end{eqnarray}
where further possible terms that are redundant owing to the  on-shell conditions 
of spin-$\frac32$ fields with $\gamma_\mu\,B_{[6]}^\mu = 0$ and $\partial_\mu\,B_{[6]}^\mu = 0$ are 
eliminated systematically.

The counter terms recalled in (\ref{del-L2S-L2V}) contribute to the baryon masses at the one-loop level. They imply renormalization scale dependent contributions that need to be balanced by a set of symmetry breaking 
counter terms in $\mathcal{L}_{\chi}^{(4)}$. We close this section with a partial collection of terms contributing to 
$\mathcal{L}_{\chi}^{(4)}$ that are relevant in a chiral extrapolation of the baryon masses at N$^3$LO. There are 16 such symmetry breaking counter terms
\begin{eqnarray}
&& \mathcal{L}_{\chi}^{(4)} \!= c_{1,[\bar 3\bar 3]}\,{\rm tr}\,\big( \bar B_{[\bar 3]}\,B_{[\bar 3]}\big)\,{\rm tr}\,\big(\chi_+^2\big)
+ c_{2,[\bar 3\bar 3]}\,{\rm tr}\,\big( \bar B_{[\bar 3]}\,B_{[\bar 3]}\big)\,\big({\rm tr}\,\chi_+\big)^2
\nonumber\\
&& \qquad \;+\, c_{3,[\bar 3\bar 3]}\,{\rm tr}\,\big( \bar B_{[\bar 3]}\,\chi_+\,B_{[\bar 3]}\,\big)\,{\rm tr}\,\big(\chi_+\big)
+  c_{4,[\bar3\bar3]}\,{\rm tr}\,\big( \bar B_{[\bar3]}\,\chi_+^2\,B_{[\bar3]}\big)
\nonumber\\
&& \qquad \; + \,c_{1,[66]}\,{\rm tr}\,\big( \bar B_{[6]}\,B_{[6]}\big)\,{\rm tr}\,\big(\chi_+^2\big)
+ c_{2,[66]}\,{\rm tr}\,\big( \bar B_{[6]}\,B_{[6]}\big)\,\big({\rm tr}\,\chi_+\big)^2
\nonumber\\
&& \qquad \;+\, c_{3,[66]}\,{\rm tr}\,\big( \bar B_{[6]}\,\chi_+\,B_{[6]}\,\big)\,{\rm tr}\,\big(\chi_+\big)
+ c_{4,[66]}\,{\rm tr}\,\big( \bar B_{[6]}\,\chi_+^2\, B_{[6]}\big)
+ c_{5,[66]}\,{\rm tr}\,\big( \bar B_{[6]}\,\chi_+\, B_{[6]}\,\chi^T_+\big)
\nonumber\\
&& \qquad \;+\, c_{1,[\bar 3 6]}\,{\rm tr}\,\big( \bar B_{[6]}\,\chi_+\,B_{[\bar 3]}+ {\rm h.c.}\,\big)\,{\rm tr}\,\big(\chi_+\big)
+  c_{2,[\bar3 6]}\,{\rm tr}\,\big( \bar B_{[6]}\,\chi_+^2\,B_{[\bar3]}+ {\rm h.c.}\big)
\nonumber\\
&& \qquad \; -\, e_{1,[66]}\,{\rm tr}\,\big( \bar B_{[6]}^\mu\,g_{\mu \nu}\,B_{[6]}^\nu\big)\,{\rm tr}\,\big(\chi_+^2\big)
- e_{2,[66]}\,{\rm tr}\,\big( \bar B_{[6]}^\mu\,g_{\mu \nu}\,B_{[6]}^\nu\big)\,\big({\rm tr}\,\chi_+\big)^2
\nonumber\\
&& \qquad \;-\, e_{3,[66]}\,{\rm tr}\,\big( \bar B_{[6]}^\mu\,g_{\mu \nu}\,\chi_+\,B_{[6]}^\nu\,\big)\,{\rm tr}\,\big(\chi_+\big)
- e_{4,[66]}\,{\rm tr}\,\big( \bar B_{[6]}^\mu\,g_{\mu \nu}\,\chi_+^2\, B_{[6]}^\nu\big)
\nonumber\\
&& \qquad \;-\, e_{5,[66]}\,{\rm tr}\,\big( \bar B_{[6]}^\mu\,g_{\mu \nu}\,\chi_+\, B_{[6]}^\nu\,\chi^T_+\big)\,.
\label{def-L4-chi}
\end{eqnarray}

Altogether we count 54 low-energy constants in this section  that have to be determined by some data set. Clearly, any additional constraints from 
heavy-quark spin symmetry or large-$N_c$ QCD are desperately needed to arrive at any significant result. Such constraints were derived in \cite{Jenkins:1996de,Lutz:2014jja,Heo:2018mur} to subleading order in the $1/N_c$ expansion and  are summarized in Appendix A for the readers' convenience.

\newpage

\section{Chiral expansion of the charmed baryon masses}

We turn to the computation of the baryon masses. The baryon self energy, $\Sigma_B(\pslash)$, may be considered to be
a function of $p_\mu \gamma^\mu $ only, with the 4-momentum $p_\mu$ of the baryon $B$. This is obvious for the spin-one-half
baryons, but less immediate for the spin-three-half baryons. We refer to \cite{Semke2005} for technical details.
To order $Q^4$ the self energy receives contributions of symmetry breaking counter terms, the tadpole  and the one-loop bubble diagram
\begin{eqnarray}
&& \Sigma_B(M_B) =  \Sigma^{\rm tree-level}_B + \Sigma^{\rm tadpole}_B + \Sigma^{\rm bubble }_B \,,
\label{def-Sigma}
\end{eqnarray}
where the index $B$ stands for the members of the flavor multiplets with $J^P = \frac{1}{2}^+, \frac{3}{2}^+$. Since the $J^P=\frac{1}{2}^+$ states come either in a flavor anti-triplet 
or a flavor sextet we discriminate those states by $B\in[\bar 3]$ and $B\in [6]$. In contrast, the $J^P=\frac{3}{2}^+$ ground states are realized only in a flavor sextet. 
In order to keep these states apart from the sextet with $J^P = \frac{1}{2}^+$ we label the $J^P = \frac{3}{2}^+$ states by $B \in [4]$ where we refer to the spin rather than the flavor 
multiplicity in this case. 

The on-shell mass of the baryon $M_B$ is determined by the condition
\begin{eqnarray}
M_B -\Sigma_B(M_B)    = M_B^{(0)} = \left\{\begin{array}{ll}
\bar M^{J= 1/2}_{[\bar 3]} \equiv M_{[\bar 3]}  & \qquad {\rm for } \qquad B\in[\bar 3]\\
\bar M^{J=1/2}_{[6]} \equiv M_{[6]} & \qquad {\rm for } \qquad B\in[6]\\
\bar M^{J=3/2}_{[6]} \equiv M_{[4]} & \qquad {\rm for }\qquad B \in[4]
\end{array} \right. \,,
\label{def-non-linear-system}
\end{eqnarray}
where $M_{[\bar 3]}, M_{[4]} $ and $ M_{[6]} $ are the renormalized and scale-independent  masses of the baryon multiplets in the flavor $SU(3)$ limit.

The separation of the baryon self energies into a loop and a tree-level contribution is not unique depending on the renormalization scheme. In this work we apply a recent
approach developed for the chiral extrapolation of the baryon octet and decuplet masses \cite{Lutz:2014oxa,Lutz:2018cqo,Guo:2018kno}. It is based on the $\chi$MS scheme \cite{Semke2005} 
and can directly be adapted to the charmed baryons, the focus of the current work. A matching with alternative renormalization schemes is most economically performed by a direct comparison
with the explicit expressions of our study. Given the renormalization scheme \cite{Lutz:2018cqo} 
the low-energy constants $b_n$ or $d_n$ do specify the linear quark mass dependence of the baryon masses as already detailed in (\ref{def-Sigma-Q2}). The particular subtraction scheme for the loop contributions as 
introduced in \cite{Lutz:2018cqo} was constructed to ensure this property of $b_n$ or $d_n$.

\begin{table}[t]
\setlength{\tabcolsep}{2.mm}
\setlength{\arraycolsep}{4.2mm}
\renewcommand{\arraystretch}{1.15}
  \centering
  \begin{tabular}{cc||c|c}
    $B$ & $Q$ & $G^{(\chi)}_{BQ}$ & $G^{(V)}_{BQ}$
    \\ \hline \hline
    $\Lambda_c$ & $\pi$ &$
    12\,B_0\,\big(2\,b_{1,[\bar{3}\bar{3}]}+b_{2,[\bar{3}\bar{3}]} \big)\,m
    $ & $
    6\,\big(g^{(V)}_{0,[\bar{3}\bar{3}]}-g^{(V)}_{1,[\bar{3}\bar{3}]}+g^{(V)}_{D,[\bar{3}\bar{3}]} \big)
    $ \\
    & $K$ &$
    4\,B_0\,\big(4\,b_{1,[\bar{3}\bar{3}]}+b_{2,[\bar{3}\bar{3}]} \big)\,(m+m_s)
    $ & $
    8\,g^{(V)}_{0,[\bar{3}\bar{3}]}+4\,g^{(V)}_{D,[\bar{3}\bar{3}]}
    $ \\
    & $\eta$ & $
    \dfrac{4}{3}\,B_0\,\big(b_{2,[\bar{3}\bar{3}]}\,m+2\,b_{1,[\bar{3}\bar{3}]}\,(m+2\,m_s) \big)
    $ & $
    2\,g^{(V)}_{0,[\bar{3}\bar{3}]} + \dfrac{2}{3}\,g^{(V)}_{1,[\bar{3}\bar{3}]} + \dfrac{2}{3}\,g^{(V)}_{D,[\bar{3}\bar{3}]}
    $
    \\
    $\Xi_c$ & $\pi$ & $
    6\,B_0\,(4\,b_{1,[\bar{3}\bar{3}]}+b_{2,[\bar{3}\bar{3}]})\,m
    $ & $
     6\,g^{(V)}_{0,[\bar{3}\bar{3}]} +3\,g^{(V)}_{D,[\bar{3}\bar{3}]}
    $ \\
    & $K$ & $
    2\,B_0\,\big(8\,b_{1,[\bar{3}\bar{3}]}+3\,b_{2,[\bar{3}\bar{3}]}\big)\,(m+m_s)
    $ & $
     8\,g^{(V)}_{0,[\bar{3}\bar{3}]}- 4\,g^{(V)}_{1,[\bar{3}\bar{3}]}+6\,g^{(V)}_{D,[\bar{3}\bar{3}]}
    $ \\
    & $\eta$ &$
    \dfrac{2}{3}\,B_0\,\big(4\,b_{1,[\bar{3}\bar{3}]}\,(m+2\,m_s)+b_{2,[\bar{3}\bar{3}]}\,(m+4\,m_s) \big)
    $ & $
    2\,g^{(V)}_{0,[\bar{3}\bar{3}]}-\dfrac{4}{3}\,g^{(V)}_{1,[\bar{3}\bar{3}] }+\dfrac{5}{3}\, \,g^{(V)}_{D,[\bar{3}\bar{3}]} 
    $
    \\ \\ \hline \hline
    
    $\Xi_c\,\Xi'_c$ & $\pi$ & $
    6\,B_0\,b_{1,[\bar{3} 6]}\,m
    $ & $
     3\,g^{(V)}_{D,[\bar{3}6]}
    $ \\
    & $K$ & $
    -2\,B_0\,b_{1,[\bar{3} 6]}\,(m+m_s)
    $ & $
    -2\,g^{(V)}_{D,[\bar{3} 6]}
    $ \\
    & $\eta$ &$
     -\dfrac{2}{3}\,B_0\,b_{1,[\bar{3} 6]}\,(4\,m_s-m)
    $ & $
    -g^{(V)}_{D,[\bar{3} 6]} 
    $
    \\ \\ \hline \hline
    
    $\Sigma_c$ & $\pi$ & $
    12\,B_0\,\big(2\,b_{1,[66]}+b_{2,[66]} \big)\,m
    $ & $
    6\,g^{(V)}_{0,[66]}+\,g^{(V)}_{1,[66]}+ 6\,g^{(V)}_{D,[66]}
    $ \\
    & $K$ & $
    4\,B_0\,\big(4\,b_{1,[66]}+b_{2,[66]} \big)\,(m+m_s)
    $ & $
    8\,g^{(V)}_{0,[66]}+4\,g^{(V)}_{D,[66]}
    $ \\
    & $\eta$ & $
    \dfrac{4}{3}\,B_0\,\big(b_{2,[66]}\,m+2\,b_{1,[66]}\,(m+2\,m_s)\big)
    $ & $
     2\,g^{(V)}_{0,[66]}+\dfrac{1}{3}\,g^{(V)}_{1,[66]}+\dfrac{2}{3}\,g^{(V)}_{D,[66]}
    $
    \\
    $\Xi'_c$ & $\pi$ & $
    6\,B_0\,\big(4\,b_{1,[66]}+b_{2,[66]}\big)\,m
    $ & $
     6\,g^{(V)}_{0,[66]}+3\,g^{(V)}_{D,[66]}
    $ \\
    & $K$ & $
    2\,B_0\,\big(8\,b_{1,[66]}+3\,b_{2,[66]}\big)\,(m+m_s)
    $ & $
    8\,g^{(V)}_{0,[66]}+2\,g^{(V)}_{1,[66]}+6\,g^{(V)}_{D,[66]}
    $ \\
    & $\eta$ & $
    \dfrac{2}{3}\,B_0\,\big(4\,b_{1,[66]}\,(m+2\,m_s)+b_{2,[66]}\,(m+4\,m_s)\big)
    $ & $
    2\,g^{(V)}_{0,[66]}-\dfrac{2}{3}\,\,g^{(V)}_{1,[66]}+\dfrac{5}{3}\,\,g^{(V)}_{D,[66]}
    $
    \\
    $\Omega_c$ & $\pi$ & $
    24\,B_0\,b_{1,[66]}\,m
    $ & $
    6\,g^{(V)}_{0,[66]}
    $ \\
    & $K$ & $
    8\,B_0\,\big(2\,b_{1,[66]}+b_{2,[66]}\big)\,(m+m_s)
    $ & $
    8\,\big(g^{(V)}_{0,[66]}+g^{(V)}_{D,[66]}\big)
    $ \\
    & $\eta$ & $
    \dfrac{8}{3}\,B_0\,\big(b_{1,[66]}\,m+2\,(b_{1,[66]}+b_{2,[66]})\,m_s \big)
    $ & $
    2\,g^{(V)}_{0,[66]}+\dfrac{4}{3}\,g^{(V)}_{1,[66]}+ \dfrac{8}{3}\,g^{(V)}_{D,[66]}
    $
    \\
    \\
\end{tabular}
  \caption{The Clebsch coefficients  $G^{(\chi)}_{QB}$ and $G^{(V)}_{QB}$ of (\ref{def-tadpole}) for the $J^P = \frac{1}{2}^+$ states. We assure that the scalar $G^{(S)}_{QB}$
  follow from the vector Clebsch by the universal replacement $g^{(V)}_{0,1,D} \to g^{(S)}_{0,1,D}$ where we use $g^{(S)}_{1, [\bar 3 \bar 3]} \equiv 0$ for notational convenience.   }
  \label{tab:1}
\end{table}
\clearpage

Let us begin with the tadpole contributions, which in a finite volume  take the following form
\begin{eqnarray}
&&  \Sigma^{\rm tadpole}_B = \frac{1}{(2\,f)^2}\sum_{Q\in [8]} \Big(  G^{(\chi )}_{BQ} \,\bar I^{(0)}_Q - m_Q^2\,G^{(S)}_{BQ} \,\bar I^{(0)}_Q -M^{(0)}_B \,G^{(V)}_{BQ} \,\bar I^{(2)}_Q \Big)\, \,,
\label{def-tadpole}
\end{eqnarray}
where the various Clebsch coefficients $G^{(\chi )}_{BQ}$ and $G^{(S)}_{BQ}, G^{(V)}_{BQ}$ are summarized in Tab. \ref{tab:1} for the $J^P = \frac{1}{2}^+$ states\footnote{For the tadpole contribution to the $\Xi_c \Xi_c'$ mixing it follows  $M_B^{(0)} = \frac{1}{2}(M_{[\bar 3]} + M_{[6]}) $.}. Note that like in our previous works \cite{Lutz:2018cqo,Guo:2018kno,Bavontaweepanya:2018yds} we use the letter $Q$ in a context specific manner. It may either denote a chiral order, or as in (\ref{def-tadpole}) or Tab. \ref{tab:1} 
if used as an index it runs over the eight Goldstone bosons properly grouped into their isospin multiplets. 
The set of finite-box scalar tadpole integrals $\bar I^{(n)}_Q$ were introduced in 
\cite{Lutz:2014oxa}. Here we recall their infinite volume limit only,
\begin{eqnarray}
&& \bar I^{(0)}_Q \to \frac{m_Q^2}{(4\,\pi)^2}\,\ln \left( \frac{m_Q^2}{\mu^2}\right)\,, \qquad \qquad \quad \bar I_Q^{(2)} \to  \frac{1}{4}\,m_Q^2\,\bar I^{(0)}_Q\,,
\label{def-barIQn}
\end{eqnarray}
with the renormalization scale $\mu$ of dimensional regularization. Explicit expressions for $\bar I_Q^{(0)}$ and  $\bar I_Q^{(2)}$  appropriate for their finite volume 
generalization are given in equations (2) and (19) of \cite{Lutz:2014oxa}.

It remains to detail the Clebsch coefficients  $G^{(\chi)}_{QB}$ and $G^{(S,V)}_{QB}$ for the $J^P = \frac{3}{2}^+$ states in the flavor sextet states. To do so it is useful to 
introduce the particular combinations  
\begin{eqnarray}
	&& \tilde g_{0,[66]}^{(S)} = {h}_{0,[66]}^{(S)} + \frac{h_{1,[66]}^{(S)}}{3},\qquad \!
	\tilde g_{1,[66]}^{(S)} = {h}_{4,[66]}^{(S)} + \frac{h_{5,[66]}^{(S)}}{3}, \qquad \!
	\tilde g_{D,[66]}^{(S)} = {h}_{2,[66]}^{(S)} +\frac{h_{3,[66]}^{(S)}}{3}, \nonumber\\
	&& \tilde g_{0,[66]}^{(V)} = {h}_{0,[66]}^{(V)} - \frac{h_{1,[66]}^{(S)}}{3\, M_{[4]}}, \! \qquad \!
	\tilde g_{1,[66]}^{(V)} = {h}_{1,[66]}^{(V)} - \frac{h_{5,[66]}^{(S)}}{3\, M_{[4]}}, \qquad \! \!
	\tilde g_{D,[66]}^{(V)} = {h}_{2,[66]}^{(V)} - \frac{h_{3,[66]}^{(S)}}{3\, M_{[4]}},
	\label{def-gtilde}
\end{eqnarray}
such that the desired Clebsch coefficients can be read off from Tab. \ref{tab:1} by replacing $g_{0,1,D}^{(S,V)} \to \tilde g_{0,1,D}^{(S,V)}$ together with $b_x \to d_x$ .

\begin{table}[t]
  \centering
  \begin{tabular}{c|ccc}   
  
    $\Sigma^{(4-\chi)}_{B}$ & $B=\Lambda_c$ & $B=\Xi_c$ & $B=\Xi_c\,\Xi_c'$
    \\ \hline \hline
    $m_\pi^4  $ & $3\,{\tilde c}_{2,[\bar{3}\bar{3}]} + 18\,{\tilde c}_{4,[\bar{3}\bar{3}]}$ & $3\,{\tilde c}_{2,[\bar{3}\bar{3}]} + 9\,{\tilde c}_{4,[\bar{3}\bar{3}]}$  & $ -3\, \tilde c_{2,[\bar 3 6]}$\\
    $m_K^4    $ & $4\,{\tilde c}_{2,[\bar{3}\bar{3}]} + 12\,{\tilde c}_{4,[\bar{3}\bar{3}]}$ & $4\,{\tilde c}_{2,[\bar{3}\bar{3}]} +18\, {\tilde c}_{4,[\bar{3}\bar{3}]}   $  &  $ 2\, \tilde c_{2,[\bar 3 6]}$\\
    $m_\eta^4 $ & ${\tilde c}_{2,[\bar{3}\bar{3}]} +2\, {\tilde c}_{4,[\bar{3}\bar{3}]}$ & ${\tilde c}_{2,[\bar{3}\bar{3}]} + 5\,{\tilde c}_{4,[\bar{3}\bar{3}]}   $  &  $ \tilde c_{2,[\bar 3 6]}$
    \\ \hline
    $B_0\,m\,m_\pi^2     $ & $18\,{\tilde c}_{3,[\bar{3}\bar{3}]}$   & $9\,{\tilde c}_{3,[\bar{3}\bar{3}]}  $ &  $-9\,{\tilde c}_{1,[\bar{3} 6]}$\\
    $B_0\,(m+m_s)\,m_K^2 $ & $6\,{\tilde c}_{3,[\bar{3}\bar{3}]}$   & $9\,{\tilde c}_{3,[\bar{3}\bar{3}]}  $ &  $3\,{\tilde c}_{1,[\bar{3} 6]}$ \\
    $B_0\,m\,m_\eta^2    $ & $2\,{\tilde c}_{3,[\bar{3}\bar{3}]}$  & ${\tilde c}_{3,[\bar{3}\bar{3}]} $ &  $-{\tilde c}_{1,[\bar{3} 6]}$ \\
    $B_0\,m_s\,m_\eta^2  $ & $0$  & $4\,{\tilde c}_{3,[\bar{3}\bar{3}]} $ &  $4\,{\tilde c}_{1,[\bar{3} 6]}$
    \\ \hline
    $B_0^2\,(2\,m^2 + m_s^2)$ & ${\tilde c}_{1,[\bar{3}\bar{3}]}$ & ${\tilde c}_{1,[\bar{3}\bar{3}]}$ & $ 0$
    \\ \\ \hline 

    $\Sigma^{(4-\chi)}_{B}$ & $B=\Sigma_c$ & $B=\Xi_c'$ & $B=\Omega_c$
    \\ \hline \hline
    $m_\pi^4  $ & $3\,{\tilde c}_{2,[66]} + 18\,{\tilde c}_{4,[66]} + 3\,{\tilde c}_{5,[66]}$   & $3\,{\tilde c}_{2,[66]} + 9\,{\tilde c}_{4,[66]}              $   & $3\,{\tilde c}_{2,[66]}                                      $ \\
    $m_K^4   $ & $4\,{\tilde c}_{2,[66]} + 12\,{\tilde c}_{4,[66]}                     $   & $4\,{\tilde c}_{2,[66]} +18\, {\tilde c}_{4,[66]} +6\, {\tilde c}_{5,[66]}$   & $4\,{\tilde c}_{2,[66]} +24\, {\tilde c}_{4,[66]}$\\
    $m_\eta^4 $ & ${\tilde c}_{2,[66]} +2\, {\tilde c}_{4,[66]} +\, {\tilde c}_{5,[66]}    $   & ${\tilde c}_{2,[66]} +5\,{\tilde c}_{4,[66]} -2\, {\tilde c}_{5,[66]}            $   & ${\tilde c}_{2,[66]} + 8\,{\tilde c}_{4,[66]} + 4\,{\tilde c}_{5,[66]}     $
    \\ \hline
    $B_0\,m\,m_\pi^2    $ & $18\,{\tilde c}_{3,[66]} $  & $9\,{\tilde c}_{3,[66]}  $ & $0 $\\
    $B_0\,(m+m_s)\,m_K^2$ & $6\,{\tilde c}_{3,[66]} $  & $9\,{\tilde c}_{3,[66]}  $ & $12\,{\tilde c}_{3,[66]}   $\\
    $B_0\,m\,m_\eta^2   $ & $2\,{\tilde c}_{3,[66]} $ & ${\tilde c}_{3,[66]} $ & $0 $\\
    $B_0\,m_s\,m_\eta^2 $ & $0  $ & $4\,{\tilde c}_{3,[66]} $ & $8\,{\tilde c}_{3,[66]} $
    \\ \hline
    $B_0^2\,(2\,m^2+ m^2_s)\, $ & ${\tilde c}_{1,[66]} $ & ${\tilde c}_{1,[66]} $ & ${\tilde c}_{1,[66]} $
  \end{tabular}
  \caption{Contributions to the baryon self energy  proportional to the product of two quark masses are expressed in terms of meson masses as to obtain renormalization scale invariant results. The original form from 
  (\ref{def-L4-chi}) is recovered in application of the Gell-Mann-Oakes-Renner relations, e.g. $m_\pi^2= 2\, B_0\, m$ and $m_K^2= B_0\,(m + m_s)$.}
  \label{tab:2}
\end{table}

Consider now the terms quadratic in the quark masses, which we denote with $\Sigma_B^{(4-\chi)}$ and 
are supposed to absorb the renormalization scale dependence of the tadpole terms $\Sigma_B^{\rm tadpole}$. They supplement the terms linear in the quark masses, which were already considered in (\ref{def-Sigma-Q2}). 
Together, both classes of terms are combined with $\Sigma_B^{\rm tree-level}$ in (\ref{def-Sigma}). 
We follow here our previous works \cite{Lutz:2014oxa,Lutz:2018cqo,Guo:2018kno} in which we keep the on-shell meson masses in the tadpole contributions. This requires to cast the terms quadratic in the quark masses into corresponding 
terms that depend on the meson masses in addition.  For that purpose we introduce particular parameter combinations
\begin{align}
  \label{res-tildec-33}
  {\tilde c}_{1,[\bar{3}\bar{3}]}
  & = \,
  {-}\frac{4}{33}\,\Big(33\,{c}_{1,[\bar{3}\bar{3}]}-45\,{c}_{2,[\bar{3}\bar{3}]}-15\,{c}_{3,[\bar{3}\bar{3}]}+11\,{c}_{4,[\bar{3}\bar{3}]}\Big)
  \,,\nonumber\\
  {\tilde c}_{2,[\bar{3}\bar{3}]}
  & = \,
  -\frac{2}{253}\,\Big(207\,{c}_{2,[\bar{3}\bar{3}]}+3\,{c}_{3,[\bar{3}\bar{3}]}-22\,{c}_{4,[\bar{3}\bar{3}]}\Big)
  \,,\nonumber\\
  {\tilde c}_{3,[\bar{3}\bar{3}]}
  & = \,
  -{\frac{1}{46}}\,\Big(9\,{c}_{3,[\bar{3}\bar{3}]}+26\,{c}_{4,[\bar{3}\bar{3}]}\Big)
  \,,\qquad
  {\tilde c}_{4,[\bar{3}\bar{3}]}
  =
  -\frac{1}{92}\,\Big(3\,{c}_{3,[\bar{3}\bar{3}]}-22\,{c}_{4,[\bar{3}\bar{3}]}\Big) \,,
 \nonumber\\ \nonumber\\ 
    \tilde{c}_{1,[\bar{3}{6}]}
  & = \,
  \frac{1}{46}\,\Big( 9\,
    {c}_{1,[\bar{3}{6}]} +26\,{c}_{2,[\bar{3}{6}]}
  \Big) \,, \qquad \qquad 
  \tilde{c}_{2,[\bar{3}{6}]}
   = \,
  \frac{3}{92}\,\Big(
   3\, {c}_{1,[\bar{3}{6}]} -22\,{c}_{2,[\bar{3}{6}]}
  \Big)
  \,,
 \end{align}
in terms of which our unambiguous results are simplified significantly. The self energies for the flavor anti-triplet states are detailed in the first part of Tab. \ref{tab:2} with the 
coupling constants ${\tilde c}_{n}$. As can be seen from Tab. \ref{tab:2} only the particular term $\tilde c_1$ 
keeps the original structure being a product of two quark masses. We identify a single parameter combination 
$\tilde c_3$  that probes the product of a quark mass with the second power of some meson mass.
The remaining parameters $\tilde c_2$ and $\tilde c_4$ select the terms involving the fourth power of a meson mass. 

Like in the previously studied cases \cite{Lutz:2018cqo,Guo:2018kno} there is a subtle issue as how to treat the flavor singlet structures proportional to $g_{0,[\bar 3\bar 3]}^{(S)} $ and $ b_{1,[\bar 3\bar 3]}$. While the first term 
stems from a chiral symmetric interaction, the second one from a structure that breaks the chiral symmetry explicitly. Nevertheless, the two terms end up with identical tadpole type contributions if the Gell-Mann-Oakes-Renner relations 
are used. The request of renormalization scale invariance implies that the  two contributions in (\ref{def-tadpole}) have to be dealt with identically, i.e. we take the replacement  
\begin{align}
	& g_{0,[\bar 3\bar 3]}^{(S)} \to g_{0,[\bar 3\bar 3]}^{(S)} -2\, b_{1,[\bar 3\bar 3]}\,,\qquad 
\end{align}
in $G^{(S)}_{BQ}$ but drop the contribution of $b_{1,[\bar 3\bar 3]}$  in $G^{(\chi)}_{BQ}$.

Analogous results can be derived for the flavor sextet states with $J^P = \frac{1}{2}^+$ and $J^P = \frac{3}{2}^+$. Here we detail our derivations for 
the $J^P = \frac{1}{2}^+$ states without loss of generality. The corresponding expressions for the $J^P = \frac{3}{2}^+$ follow upon the universal 
substitution $c_n \to e_n$. Consider the parameter combinations
\begin{align}  
  \label{res-tildec-66}
  {\tilde c}_{1,[66]}
  & = \,
  {-}\frac{4}{33}\,\Big(33\,{c}_{1,[66]}-45\,{c}_{2,[66]}-15\,{c}_{3,[66]}+11\,{c}_{4,[66]}-{c}_{5,[66]}\Big)
  \,,\nonumber\\
  {\tilde c}_{2,[66]}
  & = \,
  -\frac{2}{253}\,\Big(207\,{c}_{2,[66]}+3\,{c}_{3,[66]}-22\,{c}_{4,[66]}+20\,{c}_{5,[66]}\Big)
  \,,\nonumber\\
  {\tilde c}_{3,[66]}
  & = \,
 -{\frac{1}{46}}\,\Big(9\,{c}_{3,[66]}+26\,{c}_{4,[66]}+14\,{c}_{5,[66]}\Big)
  \,,\nonumber\\
  {\tilde c}_{4,[66]}
  & = \,
  -\frac{1}{92}\,\Big(3\,{c}_{3,[66]}-22\,{c}_{4,[66]}-26\,{c}_{5,[66]}\Big)
  \,,\qquad
  {\tilde c}_{5,[66]}
  =
  -{c}_{5,[66]}
  \,,
\end{align}
as used in the lower parts of Tab. \ref{tab:2}. We point at the one-to-one correspondence of the coefficients in (\ref{res-tildec-33}) and (\ref{res-tildec-66}) for all four terms but the $c_{5,[66]}$, which does not 
have a counter part in (\ref{res-tildec-33}). Note that here the  replacements
\begin{align}
	& g_{0,[66]}^{(S)} \to g_{0,[66]}^{(S)} -2\, b_{1,[66]} \qquad {\rm and} \qquad 
	 b_{1,[66]} \to 0\,, \nonumber
\end{align}
are required in $G^{(S)}_{BQ}$ and $G^{(\chi )}_{BQ}$ respectively.

We turn to the bubble loop contributions properly derived in the subtraction scheme \cite{Lutz:2018cqo,Guo:2018kno}.  The generic  form of the loop contributions can be taken over from our previous works \cite{Semke2005,Lutz:2014oxa,Lutz:2018cqo,Guo:2018kno}. 
Consider first the contributions to the masses of the $J^P = \frac{1}{2}^+$ states
\begin{table}[t]
  \centering
  \setlength{\tabcolsep}{3.mm}
\setlength{\arraycolsep}{2.2mm}
\renewcommand{\arraystretch}{1.25}
  \begin{tabular}{c|c|c|c}
    $\begin{array}{c@{\,=\,}l}
      G_{K \Xi_c}^{(\Lambda_c)} & -F_{[\bar{3}\bar{3}]} \\
      G_{\eta \Lambda_c}^{(\Lambda_c)} & \dfrac{1}{\sqrt{3}}\,F_{[\bar{3}\bar{3}]} \\
      G_{\pi \Sigma_c}^{(\Lambda_c)} & \sqrt{3}\, F_{[\bar{3}6]} \\ 
      G_{K \Xi_c^\prime}^{(\Lambda_c)} & F_{[\bar{3}6]} \\ \hline
      G_{\bar{K} \Xi_c}^{(\Omega_c)} & \sqrt{2}\, F_{[\bar{3}6]} \\
      G_{\bar{K} \Xi_c^\prime}^{(\Omega_c)} & \sqrt{2}\, F_{[66]} \\
      G_{\eta \Omega_c}^{(\Omega_c)} &{-} \dfrac{2}{\sqrt{3}}\, F_{[66]}
    \end{array}$
    &
    $\begin{array}{c@{\,=\,}l}
      G_{\pi \Xi_c}^{(\Xi_c)} & \dfrac{\sqrt{3}}{2}\, F_{[\bar{3}\bar{3}]} \\
      G_{\bar{K} \Lambda_c}^{(\Xi_c)} & \dfrac{1}{\sqrt{2}}\, F_{[\bar{3}\bar{3}]} \\
      G_{\eta \Xi_c}^{(\Xi_c)} &{-} \dfrac{1}{2 \sqrt{3}}\,F_{[\bar{3}\bar{3}]} \\
      G_{\pi \Xi_c^\prime}^{(\Xi_c)} & \dfrac{\sqrt{3}}{2}\, F_{[\bar{3}6]} \\
      G_{K \Omega_c}^{(\Xi_c)} & F_{[\bar{3}6]} \\
      G_{\bar{K} \Sigma_c}^{(\Xi_c)} &{-} \sqrt{\dfrac{3}{2}} \,F_{[\bar{3}6]} \\
      G_{\eta \Xi_c^\prime}^{(\Xi_c)} & \dfrac{\sqrt{3}}{2}\, F_{[\bar{3}6]}
    \end{array}$
    &
    $\begin{array}{c@{\,=\,}l}
      G_{\pi \Lambda_c}^{(\Sigma_c)} & F_{[\bar{3}6]} \\
      G_{K \Xi_c}^{(\Sigma_c)} & F_{[\bar{3}6]} \\
      G_{\pi \Sigma_c}^{(\Sigma_c)} & -\sqrt{2}\, F_{[66]} \\
      G_{K \Xi_c^\prime}^{(\Sigma_c)} & -F_{[66]} \\
      G_{\eta \Sigma_c}^{(\Sigma_c)} &\dfrac{1}{\sqrt{3}}\,F_{[66]}
    \end{array}$
    &
    $\begin{array}{c@{\,=\,}l}
      G_{\pi \Xi_c}^{(\Xi_c^\prime)} & \dfrac{\sqrt{3}}{2} \,F_{[\bar{3}6]} \\
      G_{\bar{K} \Lambda_c}^{(\Xi_c^\prime)} &{-} \dfrac{1}{\sqrt{2}} \,F_{[\bar{3}6]} \\
      G_{\eta \Xi_c}^{(\Xi_c^\prime)} & \dfrac{\sqrt{3}}{2}\, F_{[\bar{3}6]} \\
      G_{\pi \Xi_c^\prime}^{(\Xi_c^\prime)} & \dfrac{\sqrt{3}}{2} \,F_{[66]} \\
      G_{K \Omega_c}^{(\Xi_c^\prime)} & F_{[66]} \\
      G_{\bar{K} \Sigma_c}^{(\Xi_c^\prime)} & \sqrt{\dfrac{3}{2}}\, F_{[66]} \\
      G_{\eta \Xi_c^\prime}^{(\Xi_c^\prime)} &{-} \dfrac{1}{2 \sqrt{3}}\,F_{[66]}
    \end{array}$
    \\ \hline\\
    $\begin{array}{c@{\,=\,}l}
      G_{\pi \Sigma_c^{*}}^{(\Lambda_c)} & \sqrt{3}\, C_{[\bar{3}6]} \\
      G_{K \Xi_c^{*}}^{(\Lambda_c)} & C_{[\bar{3}6]} \\ \hline
      G_{\bar{K} \Xi_c^{*}}^{(\Omega_c)} & \sqrt{2} \,C_{[66]} \\
      G_{\eta \Omega_c^{*}}^{(\Omega_c)} &{-} \dfrac{2}{\sqrt{3}}\, C_{[66]}
    \end{array}$
    &
    $\begin{array}{c@{\,=\,}l}
      G_{\pi \Xi_c^{*}}^{(\Xi_c)} & \dfrac{\sqrt{3}}{2}\, C_{[\bar{3}6]} \\
      G_{K \Omega_c^{*}}^{(\Xi_c)} & C_{[\bar{3}6]} \\
      G_{\bar{K} \Sigma_c^{*}}^{(\Xi_c)} &{-} \sqrt{\dfrac{3}{2}}\, C_{[\bar{3}6]} \\
      G_{\eta \Xi_c^{*}}^{(\Xi_c)} & \dfrac{\sqrt{3}}{2}\, C_{[\bar{3}6]}
    \end{array}$
    &
    $\begin{array}{c@{\,=\,}l}
      G_{\pi \Sigma_c^{*}}^{(\Sigma_c)} & -\sqrt{2} \,C_{[66]} \\
      G_{K \Xi_c^{*}}^{(\Sigma_c)} & -C_{[66]} \\
      G_{\eta \Sigma_c^{*}}^{(\Sigma_c)} &\dfrac{1}{\sqrt{3}}\,C_{[66]}
    \end{array}$
    &
    $\begin{array}{c@{\,=\,}l}
      G_{\pi \Xi_c^{*}}^{(\Xi_c^{\prime})} & \dfrac{\sqrt{3}}{2}\, C_{[66]} \\
      G_{K \Omega_c^{*}}^{(\Xi_c^{\prime})} & C_{[66]} \\
      G_{\bar{K} \Sigma_c^{*}}^{(\Xi_c^{\prime})} & \sqrt{\dfrac{3}{2}} \,C_{[66]} \\
      G_{\eta \Xi_c^{*}}^{(\Xi_c^{\prime})} &{-} \dfrac{1}{2 \sqrt{3}}\,C_{[66]}
    \end{array}$
      \end{tabular}
  \caption{Meson-baryon coupling constants, $G^{(B)}_{QR}$, in the isospin basis. Only non-vanishing elements are shown for  the flavor anti-triplet and sextet states with $J^P = \frac{1}{2}^+$.}
  \label{tab:3}
\end{table}
\allowdisplaybreaks[1]
\begin{eqnarray}
&&\bar \Sigma^{\rm bubble}_{B \in [\bar 3,6]} = \sum_{Q\in [8], R\in [\bar 3,6]}
\left(\frac{G_{QR}^{(B)}}{2\,f} \right)^2  \Bigg\{
- \frac{(M_B+M_R)^2}{E_R+M_R}\, p^2_{QR}\,\bar I_{QR} 
\nonumber\\
&& \qquad \qquad \qquad \qquad \qquad \qquad \quad \quad { +}\,\frac{M_R^2-M_B^2}{2\,M_B}\, I^R_Q + 2\,\alpha_{QR}^{(B)}
\Bigg\}
\nonumber \\
&& \qquad  \;\;\;\,\,\,+\sum_{Q\in [8], R\in [4]}
\left(\frac{G_{QR}^{(B)}}{2\,f} \right)^2 \, \Bigg\{- \frac{2}{3}\,\frac{M_B^2}{M_R^2}\,\big(E_R+M_R\big)\,p_{QR}^{\,2}\,\bar I_{QR}
\nonumber\\
&& \qquad \qquad \qquad \qquad \qquad \qquad \quad \quad 
 +\, \frac{(M_R-M_B)\,(M_R+M_B)^3}{12\,M_B\,M^2_R}\,I^R_Q + \frac{4}{3}\,\alpha_{QR}^{(B)}
   \Bigg\}\,,
\label{result-bubble-36}
\end{eqnarray}
where we encounter the subtraction  terms $\alpha_{QR}^{(B)}$ (see (\ref{res-alpha-BQR})). 
We first recall the scalar tadpole and bubble integrals with
\begin{eqnarray}
&& I_Q^R \,= \frac{m_Q^2}{(4\,\pi)^2}\, \log \frac{m_Q^2}{M_R^2} + \Delta \bar I^{(0)}_Q\,,
\nonumber\\
&& \bar I_{Q R}= \Delta I_{QR} + \frac{1}{16\,\pi^2}
\left\{ \gamma_B^{R} - \left(\frac{1}{2}+\frac{m_Q^2-M_R^2}{2\,M_B^2}
\right)
\,\ln \left( \frac{m_Q^2}{M_R^2}\right)
\right.
\nonumber\\
&& \;\quad \;\,+\left.
\frac{p_{Q R}}{M_B}\,
\left( \ln \left(1-\frac{M_B^2-2\,p_{Q R}\,M_B}{m_Q^2+M_R^2} \right)
-\ln \left(1-\frac{M_B^2+2\,p_{Q R}\,M_B}{m_Q^2+M_R^2} \right)\right)
\right\}\;,
\nonumber\\
&& p_{Q R}^2 =
\frac{M_B^2}{4}-\frac{M_R^2+m_Q^2}{2}+\frac{(M_R^2-m_Q^2)^2}{4\,M_B^2} \,,\qquad \qquad 
E_R^2=M_R^2+p_{QR}^2 \,,
\label{def-scalar-bubble}
\end{eqnarray}
where all finite volume effects are collected into $\Delta \bar I^{(0)}_{Q}$ and $\Delta \bar I_{QR}$. For explicit expressions for the latter the reader is referred to \cite{Lutz:2014oxa}.
The sums in (\ref{result-bubble-36}) extend over the intermediate Goldstone bosons ($Q\in[8]$), the two baryon
flavor sextet states with ($R\in [6], [4]$) and one flavor anti-triplet  ($R\in[\bar 3]$). The coupling constants $G_{QR}^{(B)}$  are
determined in Tab. \ref{tab:3} and Tab. \ref{tab:4} by the parameters $F_{[ab]},D_{[ab]},C_{[ab]},H_{[ab]}$ as introduced in (\ref{def-L1}).

We note that all terms proportional to 
$m_Q^{2\,n}\,\bar I_Q$ with $n \geq 1$ are dropped in  (\ref{result-bubble-36}) as either higher order or as terms that can be absorbed into our tadpole terms. This requires to use renormalized low-energy parameters 
$\bar{g}^{(S,V)}$ in (\ref{def-tadpole}) of the following form
\begin{eqnarray}
  \label{eq:renomalized-constants-NLO}
    && {\bar g}^{(S)}_{0,[\bar{3}\bar{3}]}
     = \,
    {g}^{(S)}_{0,[\bar{3}\bar{3}]} + \frac{1}{3}\, g^{(S)}_{C,[\bar 3 6]}
    \,,\qquad \qquad 
    \bar{g}^{(S)}_{D,[\bar{3}\bar{3}]}
    =
    {g}^{(S)}_{D,[\bar{3}\bar{3}]}- g^{(S)}_{C,[\bar 3 6]}
       \,,\nonumber\\
    && \bar{g}^{(S)}_{0,[66]}
    = \,
    {g}^{(S)}_{0,[66]}
    \,,\qquad \quad 
    \bar{g}^{(S)}_{1,[66]}
    =
    {g}^{(S)}_{1,[66]} - \frac{2}{3}\,g^{(S)}_{C,[6 6]}
      \,,\qquad \quad 
    \bar{g}^{(S)}_{D,[66]}
    =
    {g}^{(S)}_{D,[66]}-  \frac{1}{3}\,g^{(S)}_{C,[6 6]}
    \,,\nonumber\\
    &&  \bar{g}^{(S)}_{D,[\bar{3}{6}]} = \,
    {g}^{(S)}_{D,[\bar{3}{6}]}
    - \frac{1}{6}\,\Big(\frac{C_{[{6}{6}]}}{C_{[\bar{3}{6}]}}\,g^{(S)}_{C,[\bar{3}{6}]}+ \frac{C_{[\bar{3}{6}]}}{C_{[{6}{6}]}}\,g^{(S)}_{C,[{6}{6}]} \Big)\,,
    \nonumber\\ \nonumber\\
    && \bar{g}^{(V)}_{0,[\bar{3}\bar{3}]}
     = \,
    {g}^{(V)}_{0,[\bar{3}\bar{3}]} +  \frac{1}{3}\, g^{(V)}_{C,[\bar 3 6]}
    \,,\qquad \quad
    \bar{g}^{(V)}_{1,[\bar{3}\bar{3}]}
    =
    {g}^{(V)}_{1,[\bar{3}\bar{3}]}
    \,,\qquad \quad
    \bar{g}^{(V)}_{D,[\bar{3}\bar{3}]}
    =
    {g}^{(V)}_{D,[\bar{3}\bar{3}]} - g^{(V)}_{C,[\bar 3 6]}
    \,,\nonumber\\
    && \bar{g}^{(V)}_{0,[66]}
     = \,
    {g}^{(V)}_{0,[66]}
    \,,\qquad \quad
    \bar{g}^{(V)}_{1,[66]}
    =
    {g}^{(V)}_{1,[66]} - \frac{2}{3}\,g^{(V)}_{C,[6 6]}
    \,,\qquad \quad
    \bar{g}^{(V)}_{D,[66]}
    =
    {g}^{(V)}_{D,[66]} -  \frac{1}{3}\,g^{(V)}_{C,[6 6]}\,,
    \nonumber\\
    && \bar{g}^{(V)}_{D,[\bar{3}{6}]}
    =
    {g}^{(V)}_{D,[\bar{3}{6}]}
    - \frac{1}{6}\,\Big(\frac{C_{[{6}{6}]}}{C_{[\bar{3}{6}]}}\,g^{(V)}_{C,[\bar{3}{6}]} +\frac{C_{[\bar{3}{6}]}}{C_{[{6}{6}]}}\, g^{(V)}_{C,[{6}{6}]} \Big)\,,
 \nonumber\\ \label{def-bargSV}\\
&& g^{(S)}_{C,[a 6]} = \frac{C^2_{[a 6]}}{2\,M_{[a]}} \,
\frac{4\,M_{[a]}^2+\Delta_{[a]}\,M_{[a]} -\Delta_{[a]}^2}{4\,(M_{[a]}+\Delta_{[a]})^2}\,,\qquad   \quad
g^{(V)}_{C,[a6 ]} = \frac{C^2_{[a 6]}}{4\,M^2_{[a]}} \,\left(\frac{M_{[a]}}{M_{[a]} + \Delta_{[a]}} \right)^2 \,,
\nonumber
  \end{eqnarray}
with the chiral limit mass spin splitting values $\Delta_{[a]} = M_{[4]}- M_{[a]}$ for $a = \bar 3, 6$. We use here the notation for the flavor SU(3) chiral limit baryon masses $M^{(0)}_B \leftrightarrow M_{[a]}$ as introduced with (\ref{def-non-linear-system}).

\begin{table}[t]
  \setlength{\tabcolsep}{2.mm}
\setlength{\arraycolsep}{1.2mm}
\renewcommand{\arraystretch}{1.25}
  \centering
  \begin{tabular}{c|c|c}
    $\begin{array}{c@{\,=\,}l}
      G_{\pi \Lambda_c}^{(\Sigma_c^{*})} & C_{[\bar{3}6]} \\
      G_{K \Xi_c}^{(\Sigma_c^{*})} & C_{[\bar{3}6]} \\
      G_{\pi \Sigma_c}^{(\Sigma_c^{*})} & -\sqrt{2} \,C_{[66]} \\
      G_{K \Xi_c^\prime}^{(\Sigma_c^{*})} & -\,C_{[66]} \\
      G_{\eta \Sigma_c}^{(\Sigma_c^{*})} & \dfrac{1}{\sqrt{3}}\,C_{[66]}
    \end{array}$
    &
    $\begin{array}{c@{\,=\,}l}
      G_{\pi \Xi_c}^{(\Xi_c^{*})} & \dfrac{\sqrt{3}}{2}\, C_{[\bar{3}6]} \\
      G_{\bar{K} \Lambda_c}^{(\Xi_c^{*})} & -\dfrac{1}{\sqrt{2}} \,C_{[\bar{3}6]} \\
      G_{\eta \Xi_c}^{(\Xi_c^{*})} & \dfrac{\sqrt{3}}{2} \,C_{[\bar{3}6]} \\
      G_{\pi \Xi_c^\prime}^{(\Xi_c^{*})} & \dfrac{\sqrt{3}}{2} \,C_{[66]} \\
      G_{K \Omega_c}^{(\Xi_c^{*})} & C_{[66]} \\
      G_{\bar{K} \Sigma_c}^{(\Xi_c^{*})} & \sqrt{\dfrac{3}{2}}\, C_{[66]} \\
      G_{\eta \Xi_c^\prime}^{(\Xi_c^{*})} &{-} \dfrac{1}{2 \sqrt{3}}\,C_{[66]}
    \end{array}$
    &
    $\begin{array}{c@{\,=\,}l}
      G_{\bar{K} \Xi_c}^{(\Omega_c^{*})} & \sqrt{2}\, C_{[\bar{3}6]} \\
      G_{\bar{K} \Xi_c^\prime}^{(\Omega_c^{*})} & \sqrt{2} \,C_{[66]} \\
      G_{\eta \Omega_c}^{(\Omega_c^{*})} &{-} \dfrac{2}{\sqrt{3}}\, C_{[66]}
    \end{array}$
    \\ \hline
    $\begin{array}{c@{\,=\,}l}
      G_{\pi \Sigma_c^{*}}^{(\Sigma_c^{*})} &- \sqrt{2}\, H_{[66]} \\
      G_{K \Xi_c^{*}}^{(\Sigma_c^{*})} & -\,H_{[66]} \\
      G_{\eta \Sigma_c^{*}}^{(\Sigma_c^{*})} & \dfrac{1}{\sqrt{3}}\,H_{[66]}
    \end{array}$
    &
    $\begin{array}{c@{\,=\,}l}
      G_{\pi \Xi_c^{*}}^{(\Xi_c^{*})} & \dfrac{\sqrt{3}}{2}\, H_{[66]} \\
      G_{K \Omega_c^{*}}^{(\Xi_c^{*})} & H_{[66]} \\
      G_{\bar{K} \Sigma_c^{*}}^{(\Xi_c^{*})} & \sqrt{\dfrac{3}{2}}\, H_{[66]} \\
      G_{\eta \Xi_c^{*}}^{(\Xi_c^{*})} &{-} \dfrac{1}{2 \sqrt{3}}\,H_{[66]}
    \end{array}$
    &
    $\begin{array}{c@{\,=\,}l}
      G_{\bar{K} \Xi_c^{*}}^{(\Omega_c^{*})} & \sqrt{2}\, H_{[66]} \\
      G_{\eta \Omega_c^{*}}^{(\Omega_c^{*})} &{-} \dfrac{2}{\sqrt{ 3}} \, H_{[66]}
    \end{array}$
  \end{tabular}
  \caption{Meson-baryon coupling constants, $G^{(B)}_{QR}$, in the isospin basis. Only non-vanishing elements are shown for  $B\in [4]$, i.e. the flavor sextet states with $J^P = \frac{3}{2}^+$.}
  \label{tab:4}
\end{table} 
 
Given our approach the scalar bubble loop function $\bar I_{QR}$ does not depend on the renormalization scale $\mu$. We point the reader at the subtraction terms $\gamma^{R}_{B}$ and $\alpha_{QR}^{(B)}$ in (\ref{result-bubble-36}). It is recalled that the subtraction 
\begin{eqnarray}
 && \gamma^{R}_{B} = -  \lim_{m, m_s\to 0}\,\frac{M_R^2-M_B^2}{M_B^2}\,\log \left|\frac{M_R^2-M_B^2}{M_R^2}\right| \,,
\label{def-gamma-BR}
 \end{eqnarray}
makes sure that the scalar bubble $\bar I_{QR}(M_B)$ will vanish in the chiral limit with $m_Q \to 0$ strictly. 
This protects the tree level slope parameters $ b_{n,[ab]}$ in (\ref{def-Sigma-Q2}) as advocated above. The additional term $\alpha^{(B)}_{QR}$ is required to protect a chiral theorem. There are non-analytic terms proportional to $m_Q^3$ that arise from the 
bubble loop contributions. Only in the presence of the subtraction terms $\alpha^{(B)}_{QR}$
they take their proper form.  From \cite{Lutz:2018cqo,Guo:2018kno} it is recalled
\begin{eqnarray}
&& \alpha^{(B)}_{QR}\, = \frac{\alpha_1\,\Delta^2}{(4\,\pi)^2} \, \Big( M_R - M_B -{\Delta_B}  \Big)\, \Big( \frac{\Delta\,\partial}{\partial\,\Delta} 
+ 1 \Big) \,\gamma_1 
 +  \frac{\Delta\, m_Q^2}{(4\,\pi)^2}\,\alpha_1\,\gamma_2 \,,
\nonumber\\
&&M = M_B^{(0)} \,,\qquad  \qquad   \Delta = M_R^{(0)}-M_B^{(0)}  \,,\qquad  \qquad 
 \gamma_1 = {\frac{2\,M+ \Delta}{2\,M}\,\log \frac{\Delta^2\,(2\,M + \Delta)^2}{(M+ \Delta)^4}} \,, \qquad \qquad 
\nonumber\\
&& \Delta_B = \Delta\,\frac{M_B} {M}\,,\qquad  \qquad
\gamma_2 = - {\frac{2\,M^2+ 2\,\Delta\,M+ \Delta^2}{2\,M\,(2\,M+ \Delta)}\,\log \frac{\Delta^2\,(2\,M + \Delta)^2}{(M+\Delta)^4} }
-\frac{M}{2\,M+\Delta} \,,
\nonumber\\
&& {\rm with} \quad \alpha_1  =\frac{(2\,M+\Delta)^4}{16\,M^2\,(M+\Delta)^2} \quad {\rm if} \quad R \in [4]  \quad {\rm but } \quad 
\alpha_1  = { \frac{(2\,M+\Delta)^2}{4\,M^2}} \quad {\rm if} \quad R \in [\bar 3,6]  \,,
\label{res-alpha-BQR}
\end{eqnarray}
where we note that the last expression for $\alpha_1$  in (\ref{res-alpha-BQR}) was not needed in \cite{Lutz:2018cqo} since there only one flavor multiplet of baryon states with $J^P=\frac{1}{2}^+$ occurs.

We close this section with the bubble loop contribution for the $J^P=\frac{3}{2}^+$ states. Again the form for the loop contributions can be inferred from our previous work \cite{Semke2005,Lutz:2018cqo}. We find
\allowdisplaybreaks[1]
\begin{eqnarray}
&&\bar \Sigma^{\rm bubble}_{B\in [4]} = \sum_{Q\in [8], R\in [\bar 3,6]}
\left(\frac{G_{QR}^{(B)}}{2\,f} \right)^2  \Bigg\{ - \frac{1}{3}\,\big( E_R +M_R\big)\,p_{QR}^{\,2}\,\bar I_{QR}
\nonumber\\
&& \qquad \qquad \qquad 
+\, \frac{(M_R-M_B)\,(M_R+M_B)^3 }{24\,M^3_B}\,I^R_Q +  \frac{2}{3}\,\alpha_{QR}^{(B)}
 \Bigg\}
\nonumber\\
&& \qquad \;\;\,\,\,+\sum_{Q\in [8], R\in [4]}
\left(\frac{G_{QR}^{(B)}}{2\,f} \right)^2 \, \Bigg\{
 -\frac{(M_B+M_R)^2}{9\,M_R^2}\,\frac{2\,E_R\,(E_R-M_R)+5\,M_R^2}{E_R+M_R}\,
p_{QR}^{\,2}\,\bar I_{QR}
\nonumber\\
&&\qquad \qquad \qquad 
+\,\frac{M_R^4+M_B^4+12\,M_R^2\,M_B^2-2\,M_R\,M_B\,(M_B^2+M_R^2)}{36\,M^3_B\,M^2_R}\,
(M^2_R-M^2_B)\,I^R_Q \Bigg\} \,,
\label{result-bubble-4}
\end{eqnarray}
with the Clebsch $G^{(B)}_{QR}$ listed in Tab. \ref{tab:4}. 
The renormalization of the coupling constants $\bar{h}^{(S,V)}$ from the bubble-loop diagram is
\begin{eqnarray}
  \label{eq:renomalized-constants-NLO}
    && \bar{h}^{(S)}_{n,[66]}
   = \,
  {h}^{(S)}_{n,[66]} \qquad \qquad \qquad {\rm for}\qquad n= 0,1,3,5
  \,,\nonumber\\
  && \bar{h}^{(S)}_{2,[66]}
  = \,
  {h}^{(S)}_{2,[66]} + \frac{1}{6}\,h^{{(S})}_{C,[\bar 3 6]} +  \frac{1}{6}\,h^{{(S})}_{C,[6 6]} 
  \, \qquad \qquad \bar{h}^{(S)}_{4,[66]}
   = \,
  {h}^{(S)}_{4,[66]} - \frac{1}{3}\,h^{{(S})}_{C,[\bar 3 6]} + \frac{1}{3}\,h^{{(S})}_{C,[6 6]} 
  \,,\nonumber\\ \nonumber\\
  && \bar{h}^{(V)}_{0,[66]}
   = \,
  {h}^{(V)}_{0,[66]}
  \,,\qquad \qquad \qquad 
   \bar{h}^{(V)}_{1,[66]}
   = \,
  {h}^{(V)}_{1,[66]}
  - \frac{H_{[66]}^2}{{ 9}\,M_{[4]}^2} 
  + \frac{1}{3}\,\Big( h^{{(V})}_{C,[\bar 3 6]} - h^{{(V})}_{C,[6 6]} \Big)
     \,,\nonumber\\
  &&\bar{h}^{(V)}_{2,[66]}
   = \,
  {h}^{(V)}_{2,[66]}
  - \frac{H_{[66]}^2}{{ 18}\,M_{[4]}^2} -   \frac{1}{6}\,\Big( h^{{(V})}_{C,[\bar 3 6]} + h^{{(V})}_{C,[6 6]} \Big)
  \,,
  \label{def-barhSV}\\ \nonumber\\
  && h^{{(S})}_{C,[a 6]} = \frac{C^2_{[a 6]}}{2\,M_{[a]}}\,  \frac{4\,M_{[a]}^3+ 5\,\Delta_{[a]}\,M_{[a]}^2 +2\,\Delta_{[a]}^2\,M_{[a]} }{4\,(M_{[a]}+\Delta_{[a]})^3} \,, \qquad \qquad
  h^{{(V})}_{C,[a 6]} = \frac{C^2_{[a 6]}}{4\,M_{[4]}\,M_{[a]}}  \frac{M_{[a]}}{M_{[a]}+\Delta_{[a]}}\,,
  \nonumber
  \end{eqnarray}
with again $\Delta_{[a]} = M_{[4]}- M_{[a]}$ for $a = \bar 3, 6$. It is left to detail 
the subtraction term $ \alpha^{(B)}_{QR}$  for the $J^P = \frac{3}{2}^+$ states which takes the form
\begin{eqnarray}
&&  \alpha^{(B)}_{QR} = \frac{\beta_1\,\Delta^2}{(4\,\pi)^2} \,\Big( M_B - M_R - {\Delta_B}\Big)\, {\Big(\frac{M+ \Delta}{M}\Big)} \Big( \frac{\Delta\,\partial}{\partial\,\Delta} + 1\Big)\,\delta_1 
 + \frac{\Delta\,m_Q^2}{(4\,\pi)^2}\,\beta_1\,\delta_2 \,, 
\nonumber\\
&&M = M_R^{(0)} \,,\qquad \qquad  \Delta = M_B^{(0)}-M_R^{(0)} \quad \qquad  \delta_1 = -\frac{M\,(2\,M+ \Delta)}{(M+\Delta)^2}\,\log \bigg|\frac{\Delta\,(2\,M+\Delta)}{M^2} \bigg|\,,
\nonumber\\
&& \Delta_B = \Delta\,\frac{M_B} {M+ \Delta}\,, \qquad \quad\delta_2 = \frac{M}{2\,M+ \Delta } +M\,\frac{2\,M^2 + 2\,\Delta\,M+\Delta^2}{(2\,M + \Delta )\,(M+ \Delta )^2}\,
\log \bigg|\frac{\Delta\,(2\,M+ \Delta)}{M^2}\bigg|\,,
\nonumber\\
&& \rm{with }\qquad  \beta_1 = \frac{(2\,M+\Delta)^4}{16\,M\,(M+\Delta)^3}\,.
\label{res-beta-BQR}
\end{eqnarray}
For a more in depth discussion of the various arguments in favour of the applied renormalization scheme  we refer to our previous works \cite{Lutz:2018cqo,Guo:2018kno}.

\section{Large-$N_c$ sum rules at the one-loop level}

In our previous work \cite{Heo:2018mur} we derived  sum rules for our low-energy constants as they arise in QCD with a large number of colors ($N_c$). The analysis was performed for tree-level expressions derived from the chiral Lagrangian. As was pointed out already for the analogous case of a study for the baryon octet and decuplet masses \cite{Lutz:2018cqo}, such relations need to be supplemented by constraints that are implied by the renormalization scale invariance condition. 

This is readily understood if one considers the scale dependence of the symmetry breaking counter terms proportional to 
$c_i$ and $e_i$ of (\ref{def-L4-chi}). The request that their contributions to the baryon masses are renormalization scale invariant is readily derived with
\begin{eqnarray}
  \mu^2\,\frac{\dd}{\dd\,\mu^2}\,c_{i,[ab]}
   = \,
  - \frac{1}{4}\,\frac{1}{(4\,\pi\,f)^2}\,\Gamma_{c_{i,[ab]}}
  \,,\qquad \qquad \quad
   \mu^2\,\frac{\dd}{\dd\,\mu^2}\,e_{i,[66]}
  = \,
  - \frac{1}{4}\,\frac{1}{(4\,\pi\,f)^2}\,\Gamma_{e_{i,[66]}}
  \,,
\label{def-Gamma}  
\end{eqnarray}
where all $\Gamma_{c_{i,[ab]}}$ and $\Gamma_{e_{i,[ab]}}$ are detailed in Appendix B. They depend on the symmetry conserving two-body terms $g$ and $h$ in(\ref{del-L2S-L2V}), but also on the symmetry breaking parameters $b$ and $d$ in  (\ref{def-L2-chi}). In turn if we insist on the leading order sum rules for the $c_i $ and $e_i$ of Appendix A the following conditions arise
\begin{eqnarray}
&&      \Gamma_{c_{n,[66]}} = \Gamma_{e_{n,[66]}}
     \qquad \qquad \qquad  \qquad \qquad \;\,\,\;\,    \Gamma_{c_{n,[\bar{3}6]}} = 0
      \qquad \quad \text{for\, all\, } n\,,
\nonumber\\
&&      \Gamma_{c_{1,[\bar{3}\bar{3}]}} = \Gamma_{c_{1,[66]}} + \Gamma_{c_{5,[66]}}/2
      \,,\qquad \qquad \qquad 
      \Gamma_{c_{2,[\bar{3}\bar{3}]}} = \Gamma_{c_{2,[66]}} - \Gamma_{c_{5,[66]}}/2
      \,,
\nonumber\\
&&      \Gamma_{c_{3,[\bar{3}\bar{3}]}} = \Gamma_{c_{3,[66]}} +2 \,\Gamma_{c_{5,[66]}}
      \,,\qquad \qquad \qquad \;\,
      \Gamma_{c_{4,[\bar{3}\bar{3}]}} = \Gamma_{c_{4,[66]}} - 2\,\Gamma_{c_{5,[66]}}
      \,.
\end{eqnarray}
If supplemented by the leading order sum rules for the remaining low-energy constants we arrive at the additional 
relations
\begin{eqnarray}
&&  \bar g^{(S)}_{1,[66]} = \,
  2\,\bar g^{(S)}_{0,[\bar{3}\bar{3}]}
  = - \frac{1}{4}\,M\,\bar g^{(V)}_{1,[66]} \qquad \!\! \!\qquad 
  \qquad \,{\rm with}
\nonumber\\
&&  M=  M_{[ 4]}= M_{[6]}= M_{[\bar 3]}\qquad \qquad {\rm or} \qquad \qquad  \bar g^{(V)}_{1,[66]} = 0 = \bar g^{(V)}_{D,[66]}\,.
\label{res-scale-LO}
\end{eqnarray}
This is an amazing prediction since now altogether we have 40 = 36 + 4 sum rules at leading order. Thus from the 54 low-energy constants we started out, there remain only  5 = 14 - 8 - 1 parameters that we have to adjust to the QCD lattice data set on the charmed baryon masses. In our parameter count we subtract the 8 charmed baryon masses known from the PDG and one axial coupling constant $C_{[\bar 3 6]}$ which is determined by the empirically known decay process $\Sigma_c^{++}(2520)\rightarrow\Lambda_c^+\pi^+$.

We close this section by a study of such sum rules at subleading order in the $1/N_c$ expansion. From Appendix A we obtain
the following conditions
\begin{eqnarray}
&& 3\,\Gamma_{c_{1,[\bar{3}\bar{3}]}}+\Gamma_{c_{2,[\bar{3}\bar{3}]}}+\Gamma_{c_{4,[\bar{3}\bar{3}]} }
   = \,
  3\,\Gamma_{c_{1,[{6}{6}]}}+\Gamma_{c_{2,[{6}{6}]}}+\Gamma_{c_{4,[{6}{6}]}}-\Gamma_{c_{5,[{6}{6}]} } 
  \,,\nonumber\\
&& 3\,\Gamma_{e_{1,[{6}{6}]}}+\Gamma_{e_{4,[{6}{6}]}}-\Gamma_{e_{5,[{6}{6}]}}
   = \,
  3\,\Gamma_ {c_{1,[{6}{6}]}}+\Gamma_{c_{4,[{6}{6}]}}-\Gamma_{c_{5,[{6}{6}]} }    \,, \qquad 
  \qquad  \Gamma_{c_{2,[66]}} = \Gamma_{e_{2,[66]}}\,,
\nonumber\\
&&      \Gamma_{c_{1,[\bar{3}6]}} =\textcolor{black}{ \big(\Gamma_{c_{3,[66]}} - \Gamma_{e_{3,[66]}} \big)/\sqrt{3} }
      \,, \qquad \qquad    \Gamma_{c_{2,[\bar{3}6]}} = \textcolor{black}{\big(\Gamma_{c_{4,[66]}} - \Gamma_{e_{4,[66]}}\big)/\sqrt{3}}
      \,,
\end{eqnarray}
which we supplement by sum rules for the low-energy constants valid at subleading order. This leads to the following five additional 
conditions
\begin{eqnarray}
&&  M_{[6]}\,\bar g^{(V)}_{0,[{6}{6}]}
   = 
 \frac{16}{39}\, \big(b_{2,[{6}{6}]}-d_{2,[{6}{6}]} \big) -4\,\bar g^{(S)}_{0,[66]} + 
4\,\bar h^{(S)}_{0,[66]} + \bar h^{(S)}_{1,[66]}+ M_{[4]}\,\bar h^{(V)}_{0 ,[66]} 
  \,,\nonumber\\
&&  \bar g^{(S)}_{D,[{6}{6}]} = 
  \bar h^{(S)}_{2,[{6}{6}]}
  + \frac{({M}_{[\bar{3}]}+{M}_{[{6}]})^2\,({M}_{[{4}]}-{M}_{[{6}]})\,({M}_{[{4}]}+{M}_{[{6}]})}{4\,{M}_{[{4}]}\,({M}_{[\bar{3}]}-{M}_{[{6}]})\,({M}_{[\bar{3}]}+3\,{M}_{[{6}]})}\,\bar h^{(V)}_{2,[{6}{6}]} 
\nonumber\\ 
&& \qquad \quad \!  -\,
\frac{3\,M^2_{[\bar{ 3}]} + 6\,M_{[\bar{3}]}\,M_{[6]} -13\,M^2_{[6]}}{8\,({M}_{[\bar{3}]}-{M}_{[{6}]})\,({M}_{[\bar{3}]}+3\,{M}_{[{6}]})}\,\bar h^{(S)}_{1,[{6}{6}]} 
  +\, \textcolor{black}{\frac{11}{13}\, \big(b_{2,[{6}{6}]}-d_{2,[{6}{6}]} \big) }  \,,
\nonumber\\
&& \bar g^{(V)}_{D,[{6}{6}]}
   = 
  \textcolor{black}{ \frac{{M}_{[{6}]}\,({M}_{[\bar{3}]}-2\,{M}_{[{4}]}+{M}_{[{6}]})\,({M}_{[\bar{3}]}+2\,{M}_{[{4}]}+{M}_{[{6}]})}{{M}_{[{4}]}\,({M}_{[\bar{3}]}-{M}_{[{6}]})\,({M}_{[\bar{3}]}+3\,{M}_{[{6}]})}\,\bar h^{(V)}_{2,[{6}{6}]} }
  \nonumber\\
 && \qquad \quad \! -\,
  \textcolor{black}{ \frac{2\,{M}_{[{6}]}\,}{({M}_{[\bar{3}]}-{M}_{[{6}]})\,({M}_{[\bar{3}]}+3\,{M}_{[{6}]})}\,\bar h^{(S)}_{1,[{6}{6}]} }  
  \,,
  \nonumber\\ 
&& \bar g^{(S)}_{0,[\bar{3}\bar{3}]} = \bar h^{(S)}_{0,[{6}{6}]} + \frac{13}{224}\,\bar h^{(S)}_{1,[{6}{6}]}
    + \frac{43}{84}\,\Big(\bar h^{(S)}_{2,[{6}{6}]} -\bar g^{(S)}_{D,[\bar{3}\bar{3}]} \Big)
    -  \frac{1}{84}\,\Big(\bar h^{(S)}_{4,[{6}{6}]}  + \frac{1}{4} \,\bar h^{(S)}_{5,[{6}{6}]}  \Big)
\nonumber\\ 
&& \qquad \quad \!  +\, \frac{{M}_{[\bar{3}]}}{4}\,\Big(- \bar g^{(V)}_{0,[\bar{3}\bar{3}]}+\frac{1}{42}\,\bar g^{(V)}_{1,[\bar{3}\bar{3}]}- \frac{43}{84}\,\bar g^{(V)}_{D,[\bar{3}\bar{3}]}\Big)
 + \frac{{M}_{[{4}]}}{4}\,\Big( \bar h^{(V)}_{0,[{6}{6}]} - \frac{1}{84}\,\bar h^{(V)}_{1,[{6}{6}]}
  +\frac{43}{84}\, \bar h^{(V)}_{2,[{6}{6}]} \Big)
  \nonumber\\
&& \qquad \quad\!   +\, \frac{15}{28}\,\big(b_{2,[\bar{3}\bar{3}]}-d_{2,[{6}{6}]}\big)
  \,,
\nonumber\\
  &&  {M}_{[{6}]}\,\bar g^{(V)}_{1,[{6}{6}]} = {M}_{[{4}]}\,\bar h^{(V)}_{1,[{6}{6}]} 
     - 4\,\bar g^{(S)}_{1,[{6}{6}]} +  4\,\bar h^{(S)}_{4,[{6}{6}]} +\bar h^{(S)}_{5,[{6}{6}]}
    \,,
  \label{res-scale-NLO}
\end{eqnarray}
where we used the three large-$N_c$ sum rules for $\bar g^{(S)}_{D,[{\bar 3 6}]}$, 
$\bar g^{(V)}_{D,[{\bar 3 6}]}$ and $\bar h^{(S)}_{3,[66]}$ only so far. 
If we use further sum rules we obtain the instrumental relation
\begin{eqnarray}
 && \bar h^{(V)}_{0,[66]} =- \frac{8\,M_{[4]}}{3}\,\frac{b_{2,[ 66]} -d_{2, [66]}}{M^2_{[4]} - M^2_{[6]}} +  \frac{1}{3}\,\Big(  \bar h^{(V)}_{1,[66]} - 2\,  \bar h^{(V)}_{2,[66]}\Big) 
 -\frac{1}{9}\,\frac{M_{[4]}}{M^2_{[4]} - M^2_{[6]}}\, \bar h_{5,[66]}^{(S)}  \,, 
\label{res-extra}
\end{eqnarray}
in terms of which all  low-energy constants can be expressed most conveniently. 
Superficially the expressions (\ref{res-extra}) appear singular at either $M_{[6]}\to M_{[4]}$ or $M_{[6]}\to M_{[\bar 3]}$. However this is not the case since in the later limits 
there are additional relations that ensure that  all low-energy constants remain finite in those limits. We remind the reader of the sum rules that arise in large-$N_c$ QCD at leading order. Here it follows $ M_{[6]} =M_{[\bar 3]} $ but also that $\bar h^{(S)}_{1,[66]}= 0$. Similarly, the heavy-quark mass limit leads to $ M_{[6]}=M_{[4]}$, but also to 
$b_{2,[66]} =d_{2,[66]}$. In turn  we may write
\begin{eqnarray}
 \bar h^{(S)}_{1,[66]} \sim \big( M_{[\bar{3}]} - M_{[6]} \big) \,, \qquad \qquad 
b_{2,[66]} - d_{2,[66]} \sim \big( M_{[6]} -M_{[4]}\big) \,.
\label{res-extra-B}
\end{eqnarray}
While we can conclude from (\ref{res-extra-B}) that the low-energy parameters in (\ref{res-scale-NLO}) remain finite at large-$N_c$ this does not yet follow in the heavy-quark mass limit. From the previous work \cite{Lutz:2014jja} we recall that there is no immediate reason that $\bar h^{(S)}_{5,[66]}$ vanishes in that limit a priori. 
Note, however, that our leading order results (\ref{res-scale-LO}) implies $\bar h^{(S)}_{5,[66]}=0$ strictly. The expressions (\ref{res-scale-NLO}) smoothly connect to our leading order findings. The first three identities in (\ref{res-scale-NLO}) approach the leading order large-$N_c$ relations $\bar g^{(V)}_{0,[66]} = 0$, $\bar g^{(S)}_{D,[{6}{6}]} = \bar h^{(S)}_{2,[{6}{6}]}$ and $\bar g^{(V)}_{D,[66]} =\bar h^{(V)}_{2,[66]} $ of Appendix A. This follows in the heavy-quark mass limit with $M_{[4]} \to M_{[6]}$ if the leading order identities $ \bar h^{(S)}_{0,[{6}{6}]} = \bar h^{(S)}_{1,[{6}{6}]} = \bar h^{(S)}_{3,[{6}{6}]} = 0 = \bar h^{(V)}_{0,[66]}$  and $\bar g^{(S)}_{0,[{6}{6}]} =0 $ are used. 
The remaining two identities recover the two scale relations $ \bar g^{(S)}_{1,[66]} = \,
  2\,\bar g^{(S)}_{0,[\bar{3}\bar{3}]}
  = - \frac{1}{4}\,M\,\bar g^{(V)}_{1,[66]}$ in this limit with $M = M_{[6]}=M_{[4]} =M_{[\bar 3]}$ (see eq. (\ref{res-scale-LO})). Note a subtle issue concerning the order at which the two limits $M_c \to \infty $ and $N_c \to  \infty$ have to be applied. Consistent results follow only if the heavy-quark mass limit with $M_{[6]}\to M_{[4]}$ at $M_{[\bar 3]}\neq M_{[4]}$ is applied first.

At subleading order altogether we have 21 = 16 + 5  sum rules. Thus from the 54 low-energy constants, there remain 
only  24 = 33 - 8 - 1 parameters that we have to adjust to the QCD lattice data set on the charmed baryon masses. Even at subleading order we deem this to be a significant result which paves the way towards a quantitative and controlled approach to chiral dynamics of charmed baryons.

\section{A convergence study for the bubble loop}

The purpose of the following section is to decompose the loop function $\bar \Sigma_B^{\rm bubble}$ into power counting moments
\begin{eqnarray}
 \bar \Sigma_B^{\rm bubble} = \bar \Sigma_B^{{\rm bubble}-3} + \bar \Sigma_B^{{\rm bubble}-4}+ \bar \Sigma_B^{{\rm bubble}-5} + \cdots \,,
 \label{def-decomposition}
\end{eqnarray}
and illustrate the convergence properties of such an expansion at hand of the physical meson and baryon masses. It is emphasized that any conventional chiral expansion in terms of bare meson and baryon masses appears futile at physical up, down and strange quark masses, at least for the baryon masses with zero charm content.  From our previous study of the chiral expansion for the charm meson masses \cite{Guo:2018kno} we already learned that such a conventional strategy appears ill defined even for charmed systems. Though, a conventional expansion for the charm baryon masses may not be as disastrous 
as it is for the baryons with zero charm content, we anticipate that our expansion in terms of on-shell masses generates much more useful and convincing results.

Given our framework and notations  the required expressions can be  readily deduced from our previous work \cite{Lutz:2018cqo}, where however a slight adaptation is necessary. Any of the moments in (\ref{def-decomposition}) receives three types of contributions
\begin{eqnarray}
 \bar \Sigma^{{\rm bubble}-n}_{B \in [a]} =  \bar \Sigma^{{\rm bubble}-n}_{B \in [a],[\bar 3]}+
  \bar \Sigma^{{\rm bubble}-n}_{B \in [a],[6]}+ \bar \Sigma^{{\rm bubble}-n}_{B \in [a],[4]}\,,
\end{eqnarray}
which are classified according to the flavor or spin multiplicity of the intermediate charmed baryon states. We will exemplify such results for the leading order term in the expansion.

For the  spin-three-half baryons in the flavor sextet we write
\begin{eqnarray}
&&  \Sigma^{{\rm bubble}-3}_{B \in [4],[4]} \!= \sum_{Q\in [8], R\in [4]} \left(\frac{1}{4\,\pi\,f}\,G_{QR}^{(B)} \right)^2 \frac{5}{9}\,
\Big\{ 
 \frac{m_Q^2}{2\,M_B}\,\Big( 1-\log\frac{m_Q}{M_R} \Big) 
 \nonumber\\
&& \qquad \qquad \qquad -\,\frac{\pi}{2}\,m_Q\Big\}\,\Big( m_Q^2- (M_R-M_B)^2\Big) \,,
\nonumber\\ \nonumber\\
&&\bar \Sigma^{{\rm bubble}-3}_{B \in [4],[\bar 3]} =\sum_{Q\in [8], R\in [\bar 3]}
\left(\frac{1}{4\,\pi\,f}\,G_{QR}^{(B)} \right)^2 \, \frac{\beta_1}{6}\,\Bigg\{ \hat \delta_2\,\Delta\,m_Q^2
\nonumber\\
&& \qquad \;\quad + \,\Big[\delta_1\,\Delta_B - \tilde \delta_1\,\big( M_B-M_R\big) \Big]\,\Delta_Q^2 -   \hat \delta_1\,\Delta^2\, \Big( M_R-M_B + \Delta_B \Big)
\nonumber\\
&& \qquad  \;\quad + \, \frac{(2\,M+ \Delta)\,M}{2\,(M+ \Delta)^2} \,
\Bigg[ \Big( \Delta_Q^2\,-  \frac{1}{2}\,m_Q^2\Big)\, \big(M_B-M_R \big)\, \log  \frac{m_Q^2}{M_R^2} 
\nonumber\\
&&  \qquad  \; \qquad \quad  +\,  \Delta_Q^3\,\Big( \log \big( M_R-M_B - \Delta_Q \big) - \log \big( M_R-M_B + \Delta_Q\big)\Big)
\Bigg]\,
\nonumber\\
&&  \qquad  \;\quad +\, \frac{m_Q^2}{\Delta_B}\, \Big( - \tilde \delta_2\,\Delta_Q^2 +\tilde \delta_3\,m_Q^2\, \log \frac{m_Q^2}{M_R^2} \Big)
\Bigg\} \,,
\nonumber\\ \nonumber\\
&&  \Delta_Q = \Big[ (M_B -M_R)^2- m_Q^2\Big]^{1/2} \,,  \qquad \qquad \qquad  \Delta_B = \Delta\,\frac{M_B}{M+ \Delta}  \,, 
\nonumber\\
&& \hat \delta_1 = \frac{(2\,M+ \Delta)}{2\,M} \,\frac{\partial }{\partial \Delta }\,\frac{2\,(M+ \Delta)}{2\,M+ \Delta}\,\Delta \,( \delta_1 - \tilde \delta_1) 
+\tilde \delta_1\,, \qquad \quad 
\nonumber\\
&& \hat \delta_2 = \delta_2 + \frac{1}{2}\,(\delta_1 - \tilde \delta_1)\,\frac{\Delta^2}{(2\,M+\Delta)^2} \,, \qquad 
\tilde \delta_1 = \delta_1 + \frac{M\,(2\,M+\Delta)}{(M + \Delta)^2}\,\ln \frac{2\,|\Delta |}{M}\,,
\label{loop-HB-3-A}
\end{eqnarray}
where $\Delta = M_{[4]}- M_{[\bar 3]}$ and $M= M_{[\bar 3]}$ in this case.  Note that the coefficients $\beta_1$ and $\delta_{1},\delta_{2}$ we encountered already in the definitions of the subtraction terms $\alpha^{(B)}_{QR}$ in (\ref{res-beta-BQR}). A complete collection of such coefficients is provided in Appendix B of \cite{Lutz:2018cqo}.

It is left to detail the contribution
$\bar \Sigma^{{\rm bubble}-3}_{B \in [4],[6]}$. It follows from $\bar \Sigma^{{\rm bubble}-3}_{B \in [4],[\bar 3]}$ by the  simple replacement ${[\bar 3] \to [6]} $ in (\ref{loop-HB-3-A}). 

We turn to the spin-one-half baryons in the flavor anti-triplet
\begin{eqnarray}
&& \bar \Sigma^{{\rm bubble}-3}_{B \in [\bar 3],[\bar 3]} \,= \sum_{Q\in [8], R\in [\bar 3]} \left(\frac{1}{4\,\pi\,f}\,G_{QR}^{(B)} \right)^2 \Big\{ 
\frac{m_Q^2}{2\,M_B}\,\Big( 1-\log\frac{m_Q}{M_R} \Big)
\nonumber\\
&& \qquad \qquad \qquad -\,\frac{\pi}{2}\,m_Q \Big\}\,\Big( m_Q^2- (M_R-M_B)^2\Big) \,,
\nonumber\\ \nonumber\\
&&\bar \Sigma^{{\rm bubble}-3}_{B \in [\bar 3],[4]} =\sum_{Q\in [8], R\in [4]}
\left(\frac{1}{4\,\pi\,f}\,G_{QR}^{(B)} \right)^2 \,\frac{\alpha_1}{3}\, \Bigg\{ \hat \gamma_2\, \Delta\,m_Q^2
\nonumber\\
&& \qquad \quad + \, \Big[\gamma_1\, \Delta_B -\tilde \gamma_1\,\big( M_R-M_B\big)\Big]\,\Delta_Q^2 + \hat \gamma_1 \,\Delta^2\,\Big( M_R-M_B -\Delta_B \Big)
\nonumber\\
&& \qquad\quad - \,\frac{2\,M+ \Delta}{2\,M}
\Bigg[ \Big( \Delta_Q^2\,-  \frac{1}{2}\,m_Q^2\Big)\,\big( M_R-M_B\big)\, \log  \frac{m_Q^2}{M_R^2} 
\nonumber\\
&& \qquad  \qquad \quad  +\,  \Delta_Q^3\,\Big( \log \big( M_R-M_B + \Delta_Q \big) - \log \big( M_R-M_B - \Delta_Q\big)\Big) \Bigg]
\nonumber\\
&&\qquad \quad +\,     \frac{m_Q^2}{\Delta_B}\, \Big[ -\tilde  \gamma_2 \,\Delta_Q^2+ \tilde \gamma_3\,m_Q^2\,\log \frac{m_Q^2}{M_R^2} \Big]\Bigg\}\,,
\nonumber\\ \nonumber\\
&&  \Delta_Q = \Big[ (M_B -M_R)^2- m_Q^2\Big]^{1/2} \,,  \qquad \qquad \qquad  \Delta_B = \Delta\,\frac{M_B}{M} \,, 
\nonumber\\
&& \hat \gamma_1 =\frac{2\,M+\Delta}{2\,M} \frac{\partial }{\partial \Delta}\,\frac{2\,\Delta\,M}{2\,M+\Delta}\,\big(\gamma_1-\tilde \gamma_1\big) + \tilde \gamma_1\,, \quad \quad 
\nonumber\\
&& \hat \gamma_2 = \gamma_2  + \frac{1}{2}\,(\gamma_1 - \tilde \gamma_1)\,\frac{\Delta^2}{(2\,M+\Delta)^2} \,,
\qquad 
\tilde \gamma_1 = \gamma_1 - \frac{2\,M+\Delta}{M }\,\ln \frac{2\,|\Delta |}{M+ \Delta}\,,
\label{loop-HB-3-B}
\end{eqnarray}
where $\Delta =M_{[4]}- M_{[\bar 3]}$ and $M= M_{[\bar 3]}$. The dimension less coefficients $\alpha_1$ and $\gamma_1, \gamma_2$ are detailed not only in (\ref{res-alpha-BQR}) but also 
in Appendix A of \cite{Lutz:2018cqo}. They depend on the ratio $\Delta/M$ only.  In this case the missing term $\bar \Sigma^{{\rm bubble}-3}_{B \in [\bar 3],[6]}$ can be obtained from $\bar \Sigma^{{\rm bubble}-3}_{B \in [\bar 3],[4]}$ with
\begin{eqnarray}
 \bar \Sigma^{{\rm bubble}-3}_{B \in [\bar 3],[6]} = \frac{3}{2}\,\bar \Sigma^{{\rm bubble}-3}_{B \in [\bar 3],[4]} \qquad 
 {\rm with}\,\quad \Delta = M_{[6]}- M_{[\bar 3]} \quad {\rm and}\quad   M = M_{[\bar 3]}\,,
\end{eqnarray}
where, however, one must use $\alpha_1 = (2\,M + \Delta)^2/(4\,M^2)$ as given already in (\ref{res-alpha-BQR}).

It remains to detail the chiral decomposition for the masses of  the spin-one-half baryons in the flavor sextet. The terms $\bar \Sigma^{{\rm bubble}-3}_{B \in [6],[6]}$ and $\bar \Sigma^{{\rm bubble}-3}_{B \in [6],[4]}$ follow directly from 
(\ref{loop-HB-3-B}) by the overall replacement $[\bar 3] \to [6]$. Then, 
the missing term $\bar \Sigma^{{\rm bubble}-3}_{B \in [6],[\bar 3]}$ is obtained from $(3/2)\,\bar \Sigma^{{\rm bubble}-3}_{B \in [6],[4]}$ by the identifications $\Delta =- M_{[6]}+ M_{[\bar 3]} < 0 $, $M = M_{[6]}$ together with  $\alpha_1 = (2\,M + \Delta)^2/(4\,M^2)$.

With the construction  of the third order terms (\ref{loop-HB-3-A}, \ref{loop-HB-3-B}) it is straightforward to correctly identify the corresponding fourth and fifth order terms from \cite{Lutz:2018cqo}. Note that the higher order terms involve additional 
coefficients $\alpha_n$, $\tilde \alpha_n$ and $\gamma_n, \tilde \gamma_n$ and 
 $\beta_n, \tilde \beta_n$ and $\delta_n, \tilde \delta_n$  that 
are detailed at the beginnings of Appendix A and B of our previous work \cite{Lutz:2018cqo}. All such coefficients are dimension less and depend on the ratio $\Delta/M$ only. It should be noted that if the ratio $\Delta/M$ turns out to be significantly smaller than $1/3$ a further 
expansion of our results in powers of such a ratio may be justified. However, this can be decided only after a full analysis of the lattice data set has been performed.  

\clearpage

\section{Some numerical results}

We now generate some numerical results illustrating the convergence properties of the chiral expansion. 
Since the relevant set of low-energy parameters is basically unknown we focus on the chiral decomposition of the one-loop bubble functions as detailed in the previous chapter. 

Any numerical estimate requires the values of the on-shell baryon and meson masses involved. Those we 
take from the PDG~\cite{Patrignani:2016xqp}. While for any of the hadron masses  $m_Q$ or $M_B, M_R$
we apply an isospin average to the values of the PDG~\cite{Patrignani:2016xqp}, for the chiral limit masses $M_{[\bar 3]}, M_{[6]}$ and $M_{[4]}$ we take the flavor SU(3) average of the corresponding multiplet masses from the PDG. The latter assumption is ad-hoc and constitutes a zeroth order estimate for such values only.  

It is left to set the axial-vector coupling constants $F$ and $C$. In \cite{Lutz:2014jja} the estimates $F_{[\bar 3 6]} \simeq 0.82$ and $C_{[\bar 3 6]} \simeq 1.36$ were derived from  
the hadronic decay widths of spin-one-half $\Sigma_c^{++}(2455)$ and spin-three-half 
 $\Sigma_c^{++}(2520)$ baryons. We provide an update of such values as is implied by the latest decay widths claimed in the PDG \cite{Patrignani:2016xqp}. We confirm that
\begin{eqnarray}
  \label{number:f36:PDG}
 && \Gamma_{\Sigma_c^{++}(2455)\rightarrow\Lambda_c^+\pi^+}
   = \,
 \phantom{0} 1.89^{+0.09}_{-0.18}\,{\rm MeV}
  \,\qquad  \rightarrow \qquad 
  |F_{[\bar{3}{6}]}| = 0.753^{+0.018}_{-0.037} \,,
 \nonumber\\
 &&  \Gamma_{\Sigma_c^{++}(2520)\rightarrow\Lambda_c^+\pi^+}
   = \,
  14.78^{+0.30}_{-0.40}\,{\rm MeV}
  \,\qquad  \rightarrow \qquad 
  |C_{[\bar{3}{6}]}| = 1.378^{+0.014}_{-0.019}\,,
  \nonumber\\
 &&  \Gamma_{\Sigma_c^{0\;}(2520)\;\,\,\rightarrow\Lambda_c^+\pi^-}
   = \,
  15.30^{+0.40}_{-0.50}\,{\rm MeV}
  \,\qquad  \rightarrow \qquad 
  |C_{[\bar{3}{6}]} |= 1.401^{+0.018}_{-0.023}
  \,,
\end{eqnarray}
translate into estimates for $ F_{[\bar{3}{6}]} $ and $ C_{[\bar{3}{6}]} $ which are compatible 
with the large-$N_c$ relation $ C_{[\bar{3}{6}]}= \sqrt{3}\,F_{[\bar{3}{6}]} $ at the 5$\%$ level. In the following we use the leading order relations (\ref{large-Nc-leading}). This leaves undetermined the axial-coupling constant $F_{[66]}$ only. 

We consider two scenarios, in the first one  we use the value $C_{[\bar{3}{6}]} = 1.35 $ together with $F_{[66]}=0$, in the second one $C_{[\bar{3}{6}]}=0$ with $F_{[66]}=1$. Once a  
value for $F_{[66]}$ is known the physical self energies can be reconstructed unambiguously in terms of our decomposition into the two cases. 
For both scenarios we illustrate with Tab. \ref{res-tab-I} and Tab. \ref{res-tab-II} that our chiral decomposition of the one-loop bubble functions is very well converging.

\begin{table}[t]
 \setlength{\tabcolsep}{4.mm}
\setlength{\arraycolsep}{1.2mm}
\renewcommand{\arraystretch}{1.25}
  \centering
  \begin{tabular}{c |c  c| c| c|c}
    B
    & $\bar \Sigma^{\rm bubble }_B$
    & $\bar \Sigma^{\rm bubble-( 3+4+5)}_B$
    & $\bar \Sigma^{\rm bubble-3}_B$
    & $\bar \Sigma^{\rm bubble-4}_B$
    & $\bar \Sigma^{\rm bubble-5}_B$
    \\ \hline \hline
    $\Lambda_c$
    & -146.20 & -146.24 & -128.46 & -17.87 & 0.09
    \\
    $\Xi_c$
    & -318.74 & -318.66 & -336.11 & 10.21 & 7.25
    \\
    $\Sigma_c$
    & -115.21 & -115.22 & -107.36 & -8.03 & 0.17
    \\
    $\Xi_c'$
    & -98.95 & -98.95 & -99.02 & -0.58 & 0.65
    \\
    $\Omega_c$
    & -94.19 & -94.18 & -97.69 & 2.87 & 0.64
    \\ \hline \hline
    $\Sigma_c^*$
    & -103.57 & -103.72 & -94.23 & -11.03 & 1.55
    \\
    $\Xi_c^*$
    & -68.87 & -68.84 & -69.27 & -2.00 & 2.43
    \\
    $\Omega_c^*$
    & -46.66 & -46.57 & -49.63 & 1.21 & 1.84
  \end{tabular}
  \caption{Baryon self energies  evaluated with physical meson and baryon masses using the leading order large-$N_c $ relations  for the axial vector coupling constants (\ref{large-Nc-leading}).  The table collects all contributions of our first scenario with $C_{[\bar 3 6]}= 1.35 =\sqrt{3}\,F_{[\bar 3 6]} $ and $ F_{[66]}= F_{[\bar 3\bar 3]}= C_{[66]}=H_{[66]}= 0$. }
  \label{res-tab-I}
\end{table}

Consider the first scenario in Tab. \ref{res-tab-I}. 
The self energy $\bar \Sigma^{\rm bubble }_B$ truncated at the fifth order is reproduced with an 
uncertainty of at most 0.2 MeV. Already with the fourth order term $\bar \Sigma^{\rm bubble-4}_B$ the full one-bubble loop function is recovered with an uncertainty of at most 7 MeV only. The contributions from the bubble loop are sizable and can be as large as 320 MeV. Thus such contributions will play a decisive role in any chiral extrapolation study of the charmed baryon masses.

\begin{table}[t]
 \setlength{\tabcolsep}{4.mm}
\setlength{\arraycolsep}{1.2mm}
\renewcommand{\arraystretch}{1.25}
  \centering
  \begin{tabular}{c| cc | c | c | c}
    B
    & $\bar \Sigma^{\rm bubble }_B$
    & $\bar \Sigma^{\rm bubble-( 3+4+5)}_B$
    & $\bar \Sigma^{\rm bubble-3}_B$
    & $\bar \Sigma^{\rm bubble-4}_B$
    & $\bar \Sigma^{\rm bubble-5}_B$
    \\ \hline \hline
     $\Lambda_c$
    & 0 & 0 & 0 & 0 & 0
    \\
    $\Xi_c$
    & 0 & 0 & 0 & 0 & 0
    \\
    $\Sigma_c$
    & -276.81 & -276.87 & -237.25 & -44.82 & 5.19
    \\
    $\Xi_c'$
    & -359.00 & -357.15 & -386.09 & 18.47 & 10.47
    \\
    $\Omega_c$
    & -473.89 & -471.62& -565.31 & 82.41 & 11.29
    \\ \hline \hline
    $\Sigma_c^*$
    & -322.17 & -322.25 & -269.53 & -60.14 & 7.42
    \\
    $\Xi_c^*$
    & -410.54 & -407.88 & -446.25 & 23.98 & 14.39
    \\
    $\Omega_c^*$
    & -535.12 & -531.82 & -653.02 & 105.61 & 15.58
  \end{tabular}
  \caption{Baryon self energies  evaluated with physical meson and baryon masses using the leading order large-$N_c $ relations  for the axial vector coupling constants (\ref{large-Nc-leading}). The table collects all contributions of our second scenario with $C_{[\bar 3 6]} = F_{[\bar 3 6]} =F_{[\bar 3 \bar 3 ]} = 0 $ and $ F_{[66]}= 1 $ together with $H_{[66]}= -\sqrt{3}\,C_{[66]} = 1.5$. }
  \label{res-tab-II}
\end{table}

We turn to our second scenario in Tab. \ref{res-tab-II}. Here we do not know the absolute size of the self energy contributions. The table shows our values at the ad-hoc choice $F_{[66]} = 1$. For instance at 
half its value with  $F_{[66]} = 0.5$ all entries in the table are reduced by a factor of four. 
Note that according to \cite{Jiang:2014ena,Yan:1992gz} the quark model suggests the value $|F_{[66]}| = 2\,|F_{[36]} |/\sqrt{3} \simeq 0.90$.  
Once a reliable estimate for the axial coupling constant $F_{[66]}$ is available the total 
contribution of the bubble loop is obtained  by adding the values in Tab. \ref{res-tab-I} with $F^2_{[66]}$
times the corresponding values of Tab. \ref{res-tab-II}. In fact such sums may be compared with the values in Tab. II of the previous work \cite{Jiang:2014ena}, which relies on the heavy-baryon mass formulation of $\chi $PT. From such a comparison we conclude again, that indeed the latter approach does not provide any significant results if truncated at N$^2$LO or N$^3$LO.

Like for the  contributions in Tab.  \ref{res-tab-I} we observe a stunning convergence behaviour. 
The self energy $\bar \Sigma^{\rm bubble }_B$ truncated at the fifth order is reproduced with an 
uncertainty of about $0.5\%$. Note that in this scenario the flavor anti-triplet baryons do not receive 
any contributions. This is so since at leading order in the large-$N_c$ expansion it holds $F_{[\bar 3 \bar 3]} =0$.

We conclude that a chiral decomposition of the one-loop contributions formulated in terms of on-shell 
meson and baryon masses appears well converging also for the charmed baryon masses. Thus, in any realistic 
application to QCD lattice data, which should be minimally at N$^3$LO, it is not required to work with loop 
expressions truncated to some order. Since the fifth order terms are about 10 MeV on average it is well justified 
to apply the loop functions as they are specified in Chapter III for the finite volume case. This is the strategy 
followed also in our previous works on the chiral extrapolation of other hadron masses in \cite{Lutz:2018cqo,Guo:2018kno}. 
Note that the size of the systemtic error in the charmed baryon masses from current QCD lattice ensembles is at least of that size.
As we repeatedly emphasized, any significant results from a fit to the lattice data can be expected only if for a given lattice ensemble 
the set of eight coupled and non-linear equations is solved that determines the charmed baryon masses.


\section{Summary}

We considered the self energies for the charmed baryon masses from the chiral Lagrangian with three light flavors at N$^3$LO. Explicit and renormalization-scale invariant expressions for all ground-state baryons with $J^P= \frac{1}{2}^+$ and $J^P= \frac{3}{2}^+$ quantum numbers  are derived. 
The results are given in terms of on-shell meson and baryon masses as it is required 
to obtain significant results that can be applied at physical up, down and strange quark masses. 
The convergence of the chiral expansion is illustrated at the hand of the one-bubble loop contributions. Given our results significant fits of the low-energy parameters to the data set on charmed baryon masses from the QCD lattice community are feasible. While at leading order in the $1/N_c$ expansion  there are 5 unknown parameters, at subleading order we derived the relevance of 21 low-energy parameters. 

\vskip0.3cm
{\bfseries{Acknowledgments}}
\vskip0.2cm
Y. Heo acknowledges partial support from Suranaree University of Technology, Office of the Higher Education Commission under NRU project of Thailand (SUT-COE: High Energy Physics and Astrophysics) and  SUT-CHE-NRU (Grant No. FtR.11/2561).

\clearpage

\section*{Appendix A}
\label{appendix:largeNc}

We provide with  Tab. \ref{tab:list-indices} a glossary of physical and technical quantities used throughout this work. Note that our notation is in part context specific. Tab. \ref{tab:LEC}
summarizes the conventions used for the various low-enery constants.

\begin{table}[b]
\begin{tabular}{c|ccc}
{\rm Index }         & {\rm values}             &   {\rm SU(3) multiplet} $\to$ {\rm label}          & {\rm defined in} \\  \hline
$Q$                  & \quad $\pi, K , \bar K, \eta $                     & $[8]\to [8]$           &  (\ref{def-states})          \\
$B, R$               & \quad $\Lambda_c$, $\Xi_c$                         & $ [\bar 3] \to [\bar 3 ] $ &   (\ref{def-states})  \\   
                     & \quad $\Xi^\mu_c$, $\Sigma^\mu_c$, $\Omega^\mu_c $ & $ [6] \to [4] $ &  (\ref{def-states}) \\ 
                     & \quad $\Xi_c'$, $\Sigma_c$, $\Omega_c $            & $ [6] \to [6] $ &  (\ref{def-states})     \\ \\

{\rm physical quantities}         & {\rm type}             &         & {\rm defined in} \\  \hline  
$m_Q $               & meson mass &   &  PDG \\
$M_B, M_R $          & baryon mass&   & PDG \\ 
$\Sigma_B $          & baryon self energy&    & (\ref{def-Sigma}, \ref{def-tadpole}, \ref{result-bubble-36}, \ref{result-bubble-4}) \\ 
$G^{(\chi)}_{BQ}$    & Clebsch coefficient &  & Tab. \ref{tab:1} \\ 
$G^{(S,V)}_{BQ}$     & Clebsch coefficient &   & Tab. \ref{tab:1} \\ 
$G^{(B)}_{QR}$       & Clebsch coefficient &  & Tab. \ref{tab:3} and Tab. \ref{tab:4}\\ \\
            
{\rm technical quantities}         & {\rm type}             &  & {\rm defined in} \\  \hline           
$\alpha^{(B)}_{QR}$  & {\rm subtraction term}              &  & (\ref{result-bubble-36}, \ref{res-alpha-BQR}, \ref{result-bubble-4}, \ref{res-beta-BQR}) \\ 
$\gamma_B^R $        & {\rm subtraction term}               &  & (\ref{def-scalar-bubble}, \ref{def-gamma-BR}) \\

$\bar I^{(n)}_Q, \bar I_Q^R$     & {\rm scalar tadpole}                 & & (\ref{def-tadpole}, \ref{def-barIQn}, \ref{def-scalar-bubble}) \\ 
$\bar I_{QR}(M_B)$        & {\rm scalar bubble}                 & & (\ref{def-scalar-bubble}) \\
$ a,b$                    & { \rm label for multiplets} $ a,b =\bar 3, 6, 4$         & & \\
$ \Gamma_{c_{n,[ab]}} $   & {\rm scale dependence}                                    & & (\ref{def-Gamma}), Appendix B \\
$ \Gamma_{e_{n,[66]}} $   & {\rm scale dependence}                                    & & (\ref{def-Gamma}), Appendix B 

\end{tabular}
 \caption{Some notations for physical and technical quantities as used in this work. }
 \label{tab:list-indices}
\end{table}

\setlength{\tabcolsep}{12.mm}
\begin{table}[t]
\begin{tabular}{c|ccc}

{\rm LEC }           &   {\rm chiral\;order}   &         & {\rm defined in} \\  \hline   
$M_{[a]}  $          &    $Q^0$                &          & (\ref{def-non-linear-system}) \\ 
$F_{[ab]} $          &     $Q^1$               &      & (\ref{def-L1}) \\
$C_{[ab]} $          &    $Q^1$                &     & (\ref{def-L1}) \\
$H_{[ab]} $          &     $Q^1$              &       & (\ref{def-L1}) \\ 

$g^{(S,V)}_{n,[ab]},\tilde g^{(S,V)}_{n,[ab]}, \bar g^{(S,V)}_{n,[ab]}$ &       $Q^2$ &                          & (\ref{del-L2S-L2V},\ref{def-gtilde}, \ref{def-bargSV}) \\

$h^{(S,V)}_{n,[ab]}, \tilde h^{(S,V)}_{n,[ab]},\bar h^{(S,V)}_{n,[ab]}$ &    $Q^2$   &                       & (\ref{del-L2S-L2V}, \ref{def-tildehSV}, \ref{def-barhSV}) \\  \hline

$b_{n,[ab]} $        &    $Q^2$    &              & (\ref{def-L2-chi}) \\
$d_{n,[ab]} $        &   $Q^2$     &                     & (\ref{def-L2-chi}) \\ 

$c_{n,[ab]}, \tilde c_{n,[ab]} $   &    $Q^4$    &    & (\ref{def-L4-chi}, \ref{res-tildec-33}, \ref{res-tildec-66}) \\
$e_{n,[ab]},\tilde e_{n,[ab]} $   &     $Q^4$    &             & (\ref{def-L4-chi}, \ref{res-tildec-66}) 
\end{tabular}
 \caption{Notation for the low-energy constants as used in this work. }
 \label{tab:LEC}
\end{table}

The large number of unknown low-energy constants is reduced by  sets of sum rules that follow from a systematic $1/N_c$ expansion \cite{Lutz:2014jja,Heo:2018mur}. While at leading order 
the large-$N_c$ operator analysis predicts $37 = 5+ 16 + 16$ sum rules
\begin{eqnarray}
&&H_{[66]} = -\sqrt 3\,C_{[66]} = \frac{3}{2}\,F_{[66]} = -
 C_{[\bar 36]} = -{\sqrt 3}\,F_{[\bar 36]}   \,, \qquad \quad 
F_{[\bar 3\bar 3]} = 0\,,
\nonumber\\ \nonumber\\
&&  b_{1,[66]} = d_{1,[66]} = b_{1,[\bar{3}\bar{3}]} \,,
  \qquad \qquad\;\;\;\; \textcolor{black}{ b_{2,[66]} = d_{2,[66]}= b_{2,[\bar{3}\bar{3}]} }
  \,,\qquad \qquad 
  b_{1,[\bar{3}6]} = 0 \,,
\nonumber\\ 
&& c_{n,[66]} = e_{n,[66]} 
\qquad \qquad \; {\rm for} \qquad n= 1,\cdots ,5 \,,\qquad
\nonumber\\
&& c_{n,[\bar 3 6]} = 0 
\qquad \qquad \qquad {\rm for} \qquad n= 1,2\,,
\nonumber\\
&& c_{1,[\bar 3\bar 3]} = c_{1,[66]} + \frac{1}{2}\,c_{5,[66]}  \qquad  \qquad
c_{2,[\bar 3\bar 3]} = c_{2,[66]}-\frac{1}{2}\,{c}_{5,[{6}{6}]}
 \,,
\nonumber\\
&& c_{3,[\bar 3\bar 3]} = c_{3,[66]}+2\,{c}_{5,[{6}{6}]} \,, \qquad  \qquad
 c_{4,[\bar 3\bar 3]} = c_{4,[66]}-2\,c_{5,[66]}  \,,
\nonumber\\ \nonumber\\
&&  {\bar g}^{(S)}_{D,[\bar{3}\bar{3}]}
   = \,
  {\bar h}^{(S)}_{2,[{6}{6}]} - {\bar h}^{(S)}_{4,[{6}{6}]} - {\bar h}^{(S)}_{5,[{6}{6}]}
  \,,\qquad \qquad 
  {\bar g}^{(S)}_{0,[\bar{3}\bar{3}]} = \frac{1}{2}\,\Big({\bar h}^{(S)}_{4,[{6}{6}]} + {\bar h}^{(S)}_{5,[{6}{6}]} \big)
  \,,\nonumber\\
&&  {\bar g}^{(S)}_{0,[{6}{6}]}
   = \,
  {\bar g}^{(S)}_{D,[\bar{3}{6}]} = 0
  \,,\qquad\qquad\qquad
  {\bar g}^{(S)}_{1,[{6}{6}]} = {\bar h}^{(S)}_{4,[{6}{6}]} + \frac{1}{3}\,{\bar h}^{(S)}_{5,[{6}{6}]}
  \,,\qquad \qquad 
  {\bar g}^{(S)}_{D,[{6}{6}]} = {\bar h}^{(S)}_{2,[{6}{6}]}
  \,,\nonumber\\
&&  {\bar h}^{(S)}_{0,[{6}{6}]}
   = \,
  {\bar h}^{(S)}_{1,[{6}{6}]} =
  {\bar h}^{(S)}_{3,[{6}{6}]} = 0
  \,,\nonumber\\
&&  \hat{g}^{(V)}_{0,[\bar{3}\bar{3}]}
   = \,
  - \hat{g}^{(V)}_{1,[\bar{3}\bar{3}]} + \frac{1}{2}\,\hat{h}^{(V)}_{1,[{6}{6}]}
  \,,\qquad \qquad \;\,
  \hat{g}^{(V)}_{D,[\bar{3}\bar{3}]}
  =
  2\,\hat{g}^{(V)}_{1,[\bar{3}\bar{3}]} - \hat{h}^{(V)}_{1,[{6}{6}]} + \hat{h}^{(V)}_{2,[{6}{6}]}
  \,,\nonumber\\
&&  \hat{g}^{(V)}_{0,[{6}{6}]}
   = \,
  \hat{g}^{(V)}_{D,[\bar{3}{6}]} =
  \hat{h}^{(V)}_{0,[{6}{6}]} = 0
  \,,\qquad \qquad 
  \hat{g}^{(V)}_{1,[{6}{6}]} = \hat{h}^{(V)}_{1,[{6}{6}]}
  \,,\qquad \qquad 
  \hat{g}^{(V)}_{D,[{6}{6}]} = \hat{h}^{(V)}_{2,[{6}{6}]}
  \,,
\label{large-Nc-leading}
\end{eqnarray}
at subleading order there remain $ 16 = 3+ 8+ 5$ sum rules only
\begin{eqnarray}
&& C_{[66]} = \textcolor{black}{ \sqrt 3\,F_{[\bar 3\bar 3]}-  \frac{1}{\sqrt 3}\,H_{[66]}} \,, \qquad \;\;\;\, F_{[\bar 3\bar 3]} = \textcolor{black}{ \frac{2}{3}\,H_{[66]} - F_{[66]} }\,,
\qquad \quad F_{[\bar 36]} = \frac{1}{\sqrt 3}\,C_{[\bar 36]}\,,
\nonumber\\ 
&& b_{1,[66]} = d_{1,[66]} = b_{1,[\bar{3}\bar{3}]}
  \,,\qquad \qquad  \quad  \;\;
  b_{1,[\bar{3}6]} = \textcolor{black}{\frac{1 }{\sqrt{3} }}\,\big(b_{2,[66]} - d_{2,[66]}\big) \,,
  \qquad \;\; 
\nonumber\\ 
&& {c}_{1,[\bar{3}{6}]}
   = \,
 \textcolor{black}{ \frac{1}{\sqrt{3}} }\,\big( {c}_{3,[{6}{6}]}-{e}_{3,[{6}{6}]}\big)
  \,,\qquad \quad 
  {c}_{2,[{6}{6}]} = {e}_{2,[{6}{6}]} \,,\qquad \quad 
{c}_{2,[\bar{3}{6}]} 
= \,  \textcolor{black}{ \frac{1}{\sqrt{3}} }\,\big( {c}_{4,[{6}{6}]}-{e}_{4,[{6}{6}]}\big)
  \,,\nonumber\\
&&  3\,{c}_{1,[\bar{3}\bar{3}]}+{c}_{2,[\bar{3}\bar{3}]}+{c}_{4,[\bar{3}\bar{3}]}
   = \,
  3\,{c}_{1,[{6}{6}]}+{c}_{2,[{6}{6}]}+{c}_{4,[{6}{6}]}-{c}_{5,[{6}{6}]}
  \,,\nonumber\\
&&  3\,{e}_{1,[{6}{6}]}+{e}_{4,[{6}{6}]}-{e}_{5,[{6}{6}]}
   = \,
  3\,{c}_{1,[{6}{6}]}+{c}_{4,[{6}{6}]}-{c}_{5,[{6}{6}]} \,,
\nonumber\\ \nonumber\\
&&  {\bar g}^{(S)}_{D,[\bar{3}{6}]}
   = \,
  \frac{1}{\sqrt{3}}\,\big({\bar g}^{(S)}_{D,[{6}{6}]} - {\bar h}^{(S)}_{2,[{6}{6}]} 
  + \frac{1}{2} \,{\bar h}^{(S)}_{1,[{6}{6}]} \big)
  \,,\qquad \qquad \qquad
 \textcolor{black}{  \bar{h}^{(S)}_{3,[{6}{6}]} = - \frac{3}{2}\,\bar{h}^{(S)}_{1,[{6}{6}]} }
  \,,\nonumber\\
&&  \textcolor{black}{  3\,\hat{g}^{(V)}_{0,[{6}{6}]} +2\,\hat{g}^{(V)}_{D,[{6}{6}]} -\hat{g}^{(V)}_{1,[{6}{6}]} = 3\,\hat{h}^{(V)}_{0,[{6}{6}]} + 2\,\hat{h}^{(V)}_{2,[{6}{6}]} -\hat{h}^{(V)}_{1,[{6}{6}]}} \,,\qquad \,
   \hat{g}^{(V)}_{D,[\bar{3}{6}]}
   = \,
  \frac{1}{\sqrt{3}}\,\big(\hat{g}^{(V)}_{D,[{6}{6}]} - \hat{h}^{(V)}_{2,[{6}{6}]}\big)
  \,,\nonumber\\
 &&  \textcolor{black}{ 3\,\bar{g}^{(S)}_{0,[{6}{6}]} +2\,\bar{g}^{(S)}_{D,[{6}{6}]} -\bar{g}^{(S)}_{1,[{6}{6}]} = 3\,\bar{h}^{(S)}_{0,[{6}{6}]} + 2\,\bar{h}^{(S)}_{2,[{6}{6}]} -\bar{h}^{(S)}_{4,[{6}{6}]} - \frac{1}{3}\,\bar{h}^{(S)}_{5,[{6}{6}]} }  
  \,,
\label{large-Nc-subleading}
\end{eqnarray}
where we correct the results for $C_{[66]}$ and $F_{[33]}$ of \cite{Lutz:2014jja} and  apply the notation 
\begin{eqnarray}
\hat g^{(V)}_{n,[ab]} = \frac{2}{M_{[a]}+ M_{[b]}}\, \bar g^{(V)}_{n,[ab]} \,,
\qquad \qquad \qquad \quad
\hat h^{(V)}_{n,[aa]} = \frac{1}{M_{[4]}}\,\bar h^{(V)}_{n,[aa]} \,.
\label{def-ghat}
\end{eqnarray}
Note that the parameter $\hat g^{(V)}_{1,[\bar 3 \bar 3]}$ cannot be determined in an analysis of the charm baryon masses. 
We close this appendix with a short summary of the implications from the heavy-quark spin symmetry that arises 
in the limit of an infinitely heavy charm quark mass \cite{Georgi:1990cx,Yan:1992gz,Cho:1992gg,Jenkins:1996de}. 
The mass parameters $M^{1/2}_{[\bar 3]}$, $M^{1/2}_{[6]}$ and $M^{3/2}_{[6]}$ may be expanded in inverse powers of the 
charm quark mass $M_c$. A matching with QCD's properties \cite{Georgi:1990cx,Yan:1992gz,Cho:1992gg,Jenkins:1996de} 
leads to the scaling properties
\begin{eqnarray}
M^{3/2}_{[6]}- M^{1/2}_{[6]} \sim \frac{1}{M_c} \,, \qquad \qquad \qquad M^{1/2}_{[6]} - M^{1/2}_{[\bar 3]} \sim M_c^0\,,
\label{larg-Mc-scalaing}
\end{eqnarray}
which implies that the two sextet masses are degenerate in this limit. We recall that the 
leading order large-$N_c$ sum rules are supplemented by one additional relation  
$ \bar h_{5,[66]}^{(S)} = 0 $ from \cite{Lutz:2014jja,Heo:2018mur} if the heavy-quark mass limit is applied. Note that at subleading order in the $1/N_c$ expansion heavy-spin symmetry breaking terms occur. For instance  $
b_{1,[\bar 36]} $ or $c_{n,[\bar 3 6]}$ need no longer to vanish.

\section*{Appendix B }

The renormalization scale dependence of the c's and e's as implied  is
\begin{eqnarray}
  \mu^2\,\frac{\dd}{\dd\,\mu^2}\,c_{i,[ab]}
   = \,
  - \frac{1}{4}\,\frac{1}{(4\,\pi\,f)^2}\,\Gamma_{c_{i,[ab]}}
  \,,\qquad \qquad 
   \mu^2\,\frac{\dd}{\dd\,\mu^2}\,e_{i,[66]}
  = \,
  - \frac{1}{4}\,\frac{1}{(4\,\pi\,f)^2}\,\Gamma_{e_{i,[66]}}
  \,,
\end{eqnarray}
with $a, b = \bar 3, 6$ and
\begin{eqnarray}
&&  \Gamma_{c_{1,[\bar{3}\bar{3}]}}
   = \,
  \frac{1}{3}\,(10\,b_{1,[\bar{3}\bar{3}]}+3\,b_{2,[\bar{3}\bar{3}]})
  - \frac{1}{9}\,(15\,\bar g^{(S)}_{0,[\bar{3}\bar{3}]}+13\,\bar g^{(S)}_{D,[\bar{3}\bar{3}]})
  \nonumber\\ 
&& \qquad \qquad  - \frac{M_{[\bar 3]}}{36}\,(15\,\bar g^{(V)}_{0,[\bar{3}\bar{3}]}-11\,\bar g^{(V)}_{1,[\bar{3}\bar{3}]}+13\,\bar g^{(V)}_{D,[\bar{3}\bar{3}]})
  \,,\nonumber\\
&&  \Gamma_{c_{2,[\bar{3}\bar{3}]}}
   = \,
  \frac{22}{9}\,b_{1,[\bar{3}\bar{3}]}
  - \frac{1}{27}\,(33\,\bar g^{(S)}_{0,[\bar{3}\bar{3}]} - 4\,\bar g^{(S)}_{D,[\bar{3}\bar{3}]})
  - \frac{M_{[\bar 3]}}{108}\,(33\,\bar g^{(V)}_{0,[\bar{3}\bar{3}]}+41\,\bar g^{(V)}_{1,[\bar{3}\bar{3}]}-4\,\bar g^{(V)}_{D,[\bar{3}\bar{3}]})
  \,,\nonumber\\
&& \Gamma_{c_{3,[\bar{3}\bar{3}]}}
   = \,
  \frac{22}{9}\,b_{2,[\bar{3}\bar{3}]}
  - \frac{26}{9}\,\bar g^{(S)}_{D,[\bar{3}\bar{3}]}
  + \frac{M_{[\bar 3]}}{36}\,(52\,\bar g^{(V)}_{1,[\bar{3}\bar{3}]}-26\,\bar g^{(V)}_{D,[\bar{3}\bar{3}]})
  \,,\nonumber\\
&& \Gamma_{c_{4,[\bar{3}\bar{3}]}}
   = \,
  \frac{1}{3}\,b_{2,[\bar{3}\bar{3}]}
  +  \bar g^{(S)}_{D,[\bar{3}\bar{3}]}
  + \frac{M_{[\bar 3]}}{4}\,(\bar g^{(V)}_{D,[\bar{3}\bar{3}]}-2\,\bar g^{(V)}_{1,[\bar{3}\bar{3}]})
  \,,\nonumber\\ \nonumber\\
&&   \Gamma_{c_{1,[\bar{3}6]}} = \,
  \frac{22}{9}\,b_{1,[\bar{3}6]}
  - \frac{26}{9}\,\bar g^{(S)}_{D,[\bar{3}6]}
  - \frac{13}{36}\,\big(M_{[\bar 3]} + M_{[6] }\big)\,\bar g^{(V)}_{D,[\bar{3}6]}
  \,,\nonumber\\
&&  \Gamma_{c_{2,[\bar{3}6]}}
   = \,
  \frac{1}{3}\,b_{1,[\bar{3}6]}
  + \bar g^{(S)}_{D,[\bar{3}6]}
  + \frac{1}{8}\,\big(M_{[\bar 3]} + M_{[6] }\big)\,\bar g^{(V)}_{D,[\bar{3}6]}
  \,,
 \nonumber\\ \nonumber\\ 
&&  \Gamma_{c_{1,[66]}}
   = \,
  \frac{1}{3}\,(10\,b_{1,[66]}+3\,b_{2,[66]})
  - \frac{1}{9}\,(15\,\bar g^{(S)}_{0,[66]}+2\,\bar g^{(S)}_{1,[66]}+13\,\bar g^{(S)}_{D,[66]})
  \nonumber\\ 
&& \qquad \qquad  - \frac{M_{[6]}}{36}\,(15\,\bar g^{(V)}_{0,[66]}+2\,\bar g^{(V)}_{1,[66]}+13\,\bar g^{(V)}_{D,[66]})
  \,,\nonumber\\
&& \Gamma_{c_{2,[66]}}
   = \,
  \frac{22}{9}\,b_{1,[66]}
  - \frac{1}{27}\,(33\,\bar g^{(S)}_{0,[66]}-2\,\bar g^{(S)}_{1,[66]}-4\,\bar g^{(S)}_{D,[66]})
  \nonumber\\ 
&& \qquad \qquad  - \frac{M_{[6]}}{108}\,(33\,\bar g^{(V)}_{0,[66]}-2\,\bar g^{(V)}_{1,[66]}-4\,\bar g^{(V)}_{D,[66]})
  \,,\nonumber\\
&& \Gamma_{c_{3,[66]}}
  = \,
  \frac{22}{9}\,b_{2,[66]}
  - \frac{2}{9}\,(2\,\bar g^{(S)}_{1,[66]}+ 13\,\bar g^{(S)}_{D,[66]})
  - \frac{M_{[6]}}{18}\,(2\,\bar g^{(V)}_{1,[66]}+13\,\bar g^{(V)}_{D,[66]})
  \,,\nonumber\\
&& \Gamma_{c_{4,[66]}}
  = \,
  \frac{1}{3}\,b_{2,[66]}
  + \frac{1}{3}\,(\bar g^{(S)}_{1,[66]}+3\,\bar g^{(S)}_{D,[66]})
  + \frac{M_{[6]}}{12}\,\Big(\bar g^{(V)}_{1,[66]}+3\,\bar g^{(V)}_{D,[66]}\Big)
  \,,\nonumber\\
&&  \Gamma_{c_{5,[66]}}
   = \,
  - \frac{1}{3}\,\bar g^{(S)}_{1,[66]}
  - \frac{M_{[6]}}{12}\,\bar g^{(V)}_{1,[66]}
  \,,\nonumber\\ \nonumber\\
&& \Gamma_{e_{1,[66]}}
  = \,
  \frac{1}{3}\,(10\,d_{1,[66]}+3\,d_{2,[66]})
  - \frac{1}{9}\,(15\,\tilde{h}^{(S)}_{0,[66]}+2\,\tilde{h}^{(S)}_{1,[66]}+13\,\tilde{h}^{(S)}_{2,[66]})
  \nonumber\\ 
&&  \qquad \qquad  - \frac{M_{[4]}}{36}\,(15\,\tilde h^{(V)}_{0,[66]}+2\,\tilde h^{(V)}_{1,[66]}+13\,\tilde h^{(V)}_{2,[66]})
  \,,\nonumber\\
&& \Gamma_{e_{2,[66]}}
   = \,
  \frac{22}{9}\,d_{1,[66]}
  - \frac{1}{27}\,(33\,\tilde{h}^{(S)}_{0,[66]}-2\,\tilde{h}^{(S)}_{1,[66]}-4\,\tilde{h}^{(S)}_{2,[66]})
  \nonumber\\ 
&& \qquad \qquad  - \frac{M_{[4]}}{108}\,(33\,\tilde h^{(V)}_{0,[66]}-2\,\tilde h^{(V)}_{1,[66]}-4\,\tilde h^{(V)}_{2,[66]})
  \,,\nonumber\\
&& \Gamma_{e_{3,[66]}}
   = \,
  \frac{22}{9}\,d_{2,[66]}
  - \frac{2}{9}\,(2\,\tilde{h}^{(S)}_{1,[66]}+13\,\tilde{h}^{(S)}_{2,[66]})
  - \frac{M_{[4]}}{18}\,(2\,\tilde h^{(V)}_{1,[66]}+13\,\tilde h^{(V)}_{2,[66]})
  \,,\nonumber\\
&& \Gamma_{e_{4,[66]}}
   = \,
  \frac{1}{3}\,d_{2,[66]}
  + \frac{1}{3}\,(\tilde{h}^{(S)}_{1,[66]}+3\,\tilde{h}^{(S)}_{2,[66]})
  + \frac{M_{[4]}}{12}\,(\tilde h^{(V)}_{1,[66]}+3\,\tilde h^{(V)}_{2,[66]})
  \,,\nonumber\\
&& \Gamma_{e_{5,[66]}}
   = \,
  - \frac{1}{3}\,\tilde{h}^{(S)}_{1,[66]}
  - \frac{M_{[4]}}{12}\,\tilde h^{(V)}_{1,[66]}
  \,,
\end{eqnarray}
and
\begin{eqnarray}
	&& \tilde h_{0,[66]}^{(S)} = {\bar h}_{0,[66]}^{(S)} + \frac{\bar h_{1,[66]}^{(S)}}{3},\qquad \!
	\tilde h_{1,[66]}^{(S)} = {\bar h}_{4,[66]}^{(S)} + \frac{\bar h_{5,[66]}^{(S)}}{3}, \qquad \!
	\tilde h_{2,[66]}^{(S)} = {\bar h}_{2,[66]}^{(S)} + \frac{\bar h_{3,[66]}^{(S)}}{3}, \nonumber\\
	&& \tilde h_{0,[66]}^{(V)} = {\bar h}_{0,[66]}^{(V)} - \frac{\bar h_{1,[66]}^{(S)}}{3\, M_{[4]} }, \! \qquad \!
	\tilde h_{1,[66]}^{(V)} = {\bar h}_{1,[66]}^{(V)} - \frac{\bar h_{5,[66]}^{(S)}}{3\, M_{[4]} }, \qquad \! \!
	\tilde h_{2,[66]}^{(V)} = {\bar h}_{2,[66]}^{(V)}- \frac{\bar h_{3,[66]}^{(S)}}{3\, M_{[4]} }\,.
\label{def-tildehSV}
\end{eqnarray}

\newpage
\bibliography{1}

\begin{thebibliography}{23}
\expandafter\ifx\csname natexlab\endcsname\relax\def\natexlab#1{#1}\fi
\expandafter\ifx\csname bibnamefont\endcsname\relax
  \def\bibnamefont#1{#1}\fi
\expandafter\ifx\csname bibfnamefont\endcsname\relax
  \def\bibfnamefont#1{#1}\fi
\expandafter\ifx\csname citenamefont\endcsname\relax
  \def\citenamefont#1{#1}\fi
\expandafter\ifx\csname url\endcsname\relax
  \def\url#1{\texttt{#1}}\fi
\expandafter\ifx\csname urlprefix\endcsname\relax\def\urlprefix{URL }\fi
\providecommand{\bibinfo}[2]{#2}
\providecommand{\eprint}[2][]{\url{#2}}

\bibitem[{\citenamefont{Lutz et~al.}(2018)\citenamefont{Lutz, Heo, and
  Guo}}]{Lutz:2018cqo}
\bibinfo{author}{\bibfnamefont{M.~F.~M.} \bibnamefont{Lutz}},
  \bibinfo{author}{\bibfnamefont{Y.}~\bibnamefont{Heo}}, \bibnamefont{and}
  \bibinfo{author}{\bibfnamefont{X.-Y.} \bibnamefont{Guo}}
  (\bibinfo{year}{2018}), \eprint{1801.06417}.

\bibitem[{\citenamefont{Guo et~al.}(2018)\citenamefont{Guo, Heo, and
  Lutz}}]{Guo:2018kno}
\bibinfo{author}{\bibfnamefont{X.-Y.} \bibnamefont{Guo}},
  \bibinfo{author}{\bibfnamefont{Y.}~\bibnamefont{Heo}}, \bibnamefont{and}
  \bibinfo{author}{\bibfnamefont{M.~F.~M.} \bibnamefont{Lutz}},
  \bibinfo{journal}{Phys. Rev.} \textbf{\bibinfo{volume}{D98}},
  \bibinfo{pages}{014510} (\bibinfo{year}{2018}).

\bibitem[{\citenamefont{Bavontaweepanya
  et~al.}(2018)\citenamefont{Bavontaweepanya, Guo, and
  Lutz}}]{Bavontaweepanya:2018yds}
\bibinfo{author}{\bibfnamefont{R.}~\bibnamefont{Bavontaweepanya}},
  \bibinfo{author}{\bibfnamefont{X.-Y.} \bibnamefont{Guo}}, \bibnamefont{and}
  \bibinfo{author}{\bibfnamefont{M.~F.~M.} \bibnamefont{Lutz}}
  (\bibinfo{year}{2018}), \eprint{1801.10522}.

\bibitem[{\citenamefont{Liu et~al.}(2010)\citenamefont{Liu, Lin, Orginos, and
  Walker-Loud}}]{Liu:2009jc}
\bibinfo{author}{\bibfnamefont{L.}~\bibnamefont{Liu}},
  \bibinfo{author}{\bibfnamefont{H.-W.} \bibnamefont{Lin}},
  \bibinfo{author}{\bibfnamefont{K.}~\bibnamefont{Orginos}}, \bibnamefont{and}
  \bibinfo{author}{\bibfnamefont{A.}~\bibnamefont{Walker-Loud}},
  \bibinfo{journal}{Phys. Rev.} \textbf{\bibinfo{volume}{D81}},
  \bibinfo{pages}{094505} (\bibinfo{year}{2010}), \eprint{0909.3294}.

\bibitem[{\citenamefont{Bali et~al.}(2013)\citenamefont{Bali, Collins, and
  Perez-Rubio}}]{Bali:2012ua}
\bibinfo{author}{\bibfnamefont{G.}~\bibnamefont{Bali}},
  \bibinfo{author}{\bibfnamefont{S.}~\bibnamefont{Collins}}, \bibnamefont{and}
  \bibinfo{author}{\bibfnamefont{P.}~\bibnamefont{Perez-Rubio}},
  \bibinfo{journal}{J.Phys.Conf.Ser.} \textbf{\bibinfo{volume}{426}},
  \bibinfo{pages}{012017} (\bibinfo{year}{2013}), \eprint{1212.0565}.

\bibitem[{\citenamefont{Briceno et~al.}(2012)\citenamefont{Briceno, Lin, and
  Bolton}}]{Briceno:2012wt}
\bibinfo{author}{\bibfnamefont{R.~A.} \bibnamefont{Briceno}},
  \bibinfo{author}{\bibfnamefont{H.-W.} \bibnamefont{Lin}}, \bibnamefont{and}
  \bibinfo{author}{\bibfnamefont{D.~R.} \bibnamefont{Bolton}},
  \bibinfo{journal}{Phys. Rev.} \textbf{\bibinfo{volume}{D86}},
  \bibinfo{pages}{094504} (\bibinfo{year}{2012}), \eprint{1207.3536}.

\bibitem[{\citenamefont{Alexandrou et~al.}(2012)\citenamefont{Alexandrou,
  Carbonell, Christaras, Drach, Gravina et~al.}}]{Alexandrou:2012xk}
\bibinfo{author}{\bibfnamefont{C.}~\bibnamefont{Alexandrou}},
  \bibinfo{author}{\bibfnamefont{J.}~\bibnamefont{Carbonell}},
  \bibinfo{author}{\bibfnamefont{D.}~\bibnamefont{Christaras}},
  \bibinfo{author}{\bibfnamefont{V.}~\bibnamefont{Drach}},
  \bibinfo{author}{\bibfnamefont{M.}~\bibnamefont{Gravina}},
  \bibnamefont{et~al.}, \bibinfo{journal}{Phys. Rev.}
  \textbf{\bibinfo{volume}{D86}}, \bibinfo{pages}{114501}
  (\bibinfo{year}{2012}), \eprint{1205.6856}.

\bibitem[{\citenamefont{Namekawa et~al.}(2013)}]{Namekawa:2013vu}
\bibinfo{author}{\bibfnamefont{Y.}~\bibnamefont{Namekawa}} \bibnamefont{et~al.}
  (\bibinfo{collaboration}{PACS-CS Collaboration}), \bibinfo{journal}{Phys.
  Rev.} \textbf{\bibinfo{volume}{D87}}, \bibinfo{pages}{094512}
  (\bibinfo{year}{2013}), \eprint{1301.4743}.

\bibitem[{\citenamefont{Perez~Rubio}(2012)}]{Perez-Rubio:2013oha}
\bibinfo{author}{\bibfnamefont{P.}~\bibnamefont{Perez~Rubio}},
  \bibinfo{journal}{PoS} \textbf{\bibinfo{volume}{ConfinementX}},
  \bibinfo{pages}{141} (\bibinfo{year}{2012}), \eprint{1302.5774}.

\bibitem[{\citenamefont{Lutz et~al.}(2014{\natexlab{a}})\citenamefont{Lutz,
  Samart, and Yan}}]{Lutz:2014jja}
\bibinfo{author}{\bibfnamefont{M.~F.~M.} \bibnamefont{Lutz}},
  \bibinfo{author}{\bibfnamefont{D.}~\bibnamefont{Samart}}, \bibnamefont{and}
  \bibinfo{author}{\bibfnamefont{Y.}~\bibnamefont{Yan}},
  \bibinfo{journal}{Phys. Rev.} \textbf{\bibinfo{volume}{D90}},
  \bibinfo{pages}{056006} (\bibinfo{year}{2014}{\natexlab{a}}),
  \eprint{1402.6427}.

\bibitem[{\citenamefont{Heo and Lutz}(2018)}]{Heo:2018mur}
\bibinfo{author}{\bibfnamefont{Y.}~\bibnamefont{Heo}} \bibnamefont{and}
  \bibinfo{author}{\bibfnamefont{M.~F.~M.} \bibnamefont{Lutz}},
  \bibinfo{journal}{Phys. Rev.} \textbf{\bibinfo{volume}{D97}},
  \bibinfo{pages}{094004} (\bibinfo{year}{2018}), \eprint{1802.09365}.

\bibitem[{\citenamefont{Jiang et~al.}(2014)\citenamefont{Jiang, Chen, and
  Zhu}}]{Jiang:2014ena}
\bibinfo{author}{\bibfnamefont{N.}~\bibnamefont{Jiang}},
  \bibinfo{author}{\bibfnamefont{X.-L.} \bibnamefont{Chen}}, \bibnamefont{and}
  \bibinfo{author}{\bibfnamefont{S.-L.} \bibnamefont{Zhu}},
  \bibinfo{journal}{Phys. Rev.} \textbf{\bibinfo{volume}{D90}},
  \bibinfo{pages}{074011} (\bibinfo{year}{2014}), \eprint{1403.5404}.

\bibitem[{\citenamefont{Yan et~al.}(1992)\citenamefont{Yan, Cheng, Cheung, Lin,
  Lin et~al.}}]{Yan:1992gz}
\bibinfo{author}{\bibfnamefont{T.-M.} \bibnamefont{Yan}},
  \bibinfo{author}{\bibfnamefont{H.-Y.} \bibnamefont{Cheng}},
  \bibinfo{author}{\bibfnamefont{C.-Y.} \bibnamefont{Cheung}},
  \bibinfo{author}{\bibfnamefont{G.-L.} \bibnamefont{Lin}},
  \bibinfo{author}{\bibfnamefont{Y.}~\bibnamefont{Lin}}, \bibnamefont{et~al.},
  \bibinfo{journal}{Phys. Rev.} \textbf{\bibinfo{volume}{D46}},
  \bibinfo{pages}{1148} (\bibinfo{year}{1992}).

\bibitem[{\citenamefont{Cho}(1992)}]{Cho:1992gg}
\bibinfo{author}{\bibfnamefont{P.~L.} \bibnamefont{Cho}},
  \bibinfo{journal}{Phys. Lett.} \textbf{\bibinfo{volume}{B285}},
  \bibinfo{pages}{145} (\bibinfo{year}{1992}), \eprint{hep-ph/9203225}.

\bibitem[{\citenamefont{Terschlüsen et~al.}(2012)\citenamefont{Terschlüsen,
  Leupold, and Lutz}}]{Terschlusen:2012xw}
\bibinfo{author}{\bibfnamefont{C.}~\bibnamefont{Terschlüsen}},
  \bibinfo{author}{\bibfnamefont{S.}~\bibnamefont{Leupold}}, \bibnamefont{and}
  \bibinfo{author}{\bibfnamefont{M.~F.~M.} \bibnamefont{Lutz}},
  \bibinfo{journal}{Eur. Phys. J.} \textbf{\bibinfo{volume}{A48}},
  \bibinfo{pages}{190} (\bibinfo{year}{2012}), \eprint{1204.4125}.

\bibitem[{\citenamefont{Guo et~al.}(2015)\citenamefont{Guo, Guo, Oller, and
  Sanz-Cillero}}]{Guo:2015xva}
\bibinfo{author}{\bibfnamefont{X.-K.} \bibnamefont{Guo}},
  \bibinfo{author}{\bibfnamefont{Z.-H.} \bibnamefont{Guo}},
  \bibinfo{author}{\bibfnamefont{J.~A.} \bibnamefont{Oller}}, \bibnamefont{and}
  \bibinfo{author}{\bibfnamefont{J.~J.} \bibnamefont{Sanz-Cillero}},
  \bibinfo{journal}{JHEP} \textbf{\bibinfo{volume}{06}}, \bibinfo{pages}{175}
  (\bibinfo{year}{2015}), \eprint{1503.02248}.

\bibitem[{\citenamefont{Okubo}(1963)}]{Okubo:1963fa}
\bibinfo{author}{\bibfnamefont{S.}~\bibnamefont{Okubo}},
  \bibinfo{journal}{Phys. Lett.} \textbf{\bibinfo{volume}{5}},
  \bibinfo{pages}{165} (\bibinfo{year}{1963}).

\bibitem[{\citenamefont{F.~M.~Lutz and Kolomeitsev}(2004)}]{Lutz:2003jw}
\bibinfo{author}{\bibfnamefont{M.}~\bibnamefont{F.~M.~Lutz}} \bibnamefont{and}
  \bibinfo{author}{\bibfnamefont{E.}~\bibnamefont{Kolomeitsev}},
  \bibinfo{journal}{Nucl.Phys.} \textbf{\bibinfo{volume}{A730}},
  \bibinfo{pages}{110} (\bibinfo{year}{2004}), \eprint{hep-ph/0307233}.

\bibitem[{\citenamefont{Jenkins}(1996)}]{Jenkins:1996de}
\bibinfo{author}{\bibfnamefont{E.~E.} \bibnamefont{Jenkins}},
  \bibinfo{journal}{Phys. Rev.} \textbf{\bibinfo{volume}{D54}},
  \bibinfo{pages}{4515} (\bibinfo{year}{1996}), \eprint{hep-ph/9603449}.

\bibitem[{\citenamefont{Semke and F.~M.~Lutz}(2006)}]{Semke2005}
\bibinfo{author}{\bibfnamefont{A.}~\bibnamefont{Semke}} \bibnamefont{and}
  \bibinfo{author}{\bibfnamefont{M.}~\bibnamefont{F.~M.~Lutz}},
  \bibinfo{journal}{Nucl. Phys.} \textbf{\bibinfo{volume}{A778}},
  \bibinfo{pages}{153} (\bibinfo{year}{2006}), \eprint{nucl-th/0511061}.

\bibitem[{\citenamefont{Lutz et~al.}(2014{\natexlab{b}})\citenamefont{Lutz,
  Bavontaweepanya, Kobdaj, and Schwarz}}]{Lutz:2014oxa}
\bibinfo{author}{\bibfnamefont{M.~F.~M.} \bibnamefont{Lutz}},
  \bibinfo{author}{\bibfnamefont{R.}~\bibnamefont{Bavontaweepanya}},
  \bibinfo{author}{\bibfnamefont{C.}~\bibnamefont{Kobdaj}}, \bibnamefont{and}
  \bibinfo{author}{\bibfnamefont{K.}~\bibnamefont{Schwarz}},
  \bibinfo{journal}{Phys. Rev.} \textbf{\bibinfo{volume}{D90}},
  \bibinfo{pages}{054505} (\bibinfo{year}{2014}{\natexlab{b}}),
  \eprint{1401.7805}.

\bibitem[{\citenamefont{Patrignani et~al.}(2016)}]{Patrignani:2016xqp}
\bibinfo{author}{\bibfnamefont{C.}~\bibnamefont{Patrignani}}
  \bibnamefont{et~al.} (\bibinfo{collaboration}{Particle Data Group}),
  \bibinfo{journal}{Chin. Phys.} \textbf{\bibinfo{volume}{C40}},
  \bibinfo{pages}{100001} (\bibinfo{year}{2016}).

\bibitem[{\citenamefont{Georgi}(1991)}]{Georgi:1990cx}
\bibinfo{author}{\bibfnamefont{H.}~\bibnamefont{Georgi}},
  \bibinfo{journal}{Nucl. Phys.} \textbf{\bibinfo{volume}{B348}},
  \bibinfo{pages}{293} (\bibinfo{year}{1991}).

\end{thebibliography}
\end{document}